**ELEVATE-GenAI: Reporting Guidelines for the Use of Large Language Models in Health**

**Economics and Outcomes Research: an ISPOR Working Group on Generative AI Report**

**Authors:** Rachael L. Fleurence, PhD[1,2], Dalia Dawoud, PhD[3,4] Jiang Bian, PhD[5,6,7] Mitchell K. Higashi, PhD[8,] Xiaoyan Wang, PhD[9,10] , Hua Xu, PhD[11], Jagpreet Chhatwal, PhD[12,13], Turgay Ayer, PhD[14,15] on behalf of The ISPOR Working Group on Generative AI.

1. Value Analytics Labs, Cambridge, MA, United States
2. Office of the Director, National Institutes of Health, National Institute of Biomedical Imaging and Bioengineering, Bethesda, MD, United States
3. National Institute for Health and Care Excellence, London, United Kingdom.
4. Cairo University, Faculty of Pharmacy, Cairo, Egypt
5. Health Outcomes and Biomedical Informatics, College of Medicine, University of Florida, FL, United States
6. Biomedical Informatics, Clinical and Translational Science Institute, University of Florida, FL, United States
7. Office of Data Science and Research Implementation, University of Florida Health, Gainesville, FL, United States
8. ISPOR, The Professional Society for Health Economics and Outcomes Research, Lawrenceville, NJ, United States
9. Tulane University School of Public Health and Tropical Medicine, New Orleans, LA
10. Intelligent Medical Objects, Rosemont, IL, United States
11. Institute Department of Biomedical Informatics and Data Science, School of Medicine, Yale University, New Haven, CT, United States
12. Institute for Technology Assessment, Massachusetts General Hospital, Harvard Medical School, Boston, MA, United States
13. Center for Health Decision Science, Harvard University, Boston, MA, United States
14. Value Analytics Labs, Cambridge, MA, United States
15. Center for Health & Humanitarian Systems, Georgia Institute of Technology, Atlanta, GA, United States

**Funding:** Dr Dalia Dawoud reports partial funding from the European Union's Horizon 2020 research and innovation programme under Grant Agreement No. 82516 (Next Generation Health Technology Assessment (HTx) project. No other funding was received.

**Acknowledgements**: This manuscript was developed as part of the International Society for Pharmacoeconomics and Outcomes Research (ISPOR) Working Group on Generative AI. The authors wish to thank the ISPOR Science office for their support, Sahar Alam for her excellent program management throughout the project. The views expressed are those of the authors and do not necessarily reflect the official policy or position of their employers, former employers, or funding organizations.

**Highlights**

**What methods or evidence gap does your paper address?**

This paper addresses the lack of structured guidance for reporting research using large language models (LLMs) in Health Economics and Outcomes Research (HEOR) by introducing the ELEVATE-GenAI framework and checklist.

**What are the key findings from your research?**

The ELEVATE-GenAI framework and checklist provides a practical, domain-specific tool for systematically reporting the use of LLMs in HEOR research, emphasizing 10 domains including transparency, accuracy, and reproducibility.

**What are the implications of your findings for healthcare decision-making or the practice of HEOR?**

The reporting guidelines promote rigorous reporting standards, enabling HEOR professionals to integrate LLMs responsibly, enhancing evidence synthesis, modeling, and real-world data generation in healthcare research.

**Abstract**

**Introduction:** Generative artificial intelligence (AI), particularly large language models (LLMs), holds significant promise for Health Economics and Outcomes Research (HEOR). However, standardized reporting guidance for LLM-assisted research is lacking. This article introduces the ELEVATE-GenAI framework and checklist—reporting guidelines specifically designed for HEOR studies involving LLMs.

**Methods:** The framework was developed through a targeted literature review of existing reporting guidelines, AI evaluation frameworks, and expert input from the ISPOR Working Group on Generative AI. It comprises ten domains—including model characteristics, accuracy, reproducibility, and fairness and bias. The accompanying checklist translates the framework into actionable reporting items. To illustrate its use, the framework was applied to two published HEOR studies: one focused on a systematic literature review tasks and the other on economic modeling.

**Results:** The ELEVATE-GenAI framework offers a comprehensive structure for reporting LLM-assisted HEOR research, while the checklist facilitates practical implementation. Its application to the two case studies demonstrates its relevance and usability across different HEOR contexts.

**Limitations:** Although the framework provides robust reporting guidance, further empirical testing is needed to assess its validity, completeness, usability as well as its generalizability across diverse HEOR use cases.

**Conclusion:** The ELEVATE-GenAI framework and checklist address a critical gap by offering structured guidance for transparent, accurate, and reproducible reporting of LLM-assisted HEOR research. Future work will focus on extensive testing and validation to support broader adoption and refinement.

**ELEVATE-GenAI: Reporting Guidelines for the Use of Large Language Models in Health Economics and Outcomes Research: an ISPOR Working Group on Generative AI Report**

**Introduction**

Artificial intelligence (AI) encompasses computational methods for tasks requiring human-like reasoning, learning, or decision-making[1]. Natural language processing (NLP), a subfield of AI, enables machines to understand and generate human language[2]. Generative AI models produce new content—such as text, code, or data—based on patterns in training data[3,4], with large language models (LLMs) emerging as especially impactful. Foundation models like GPT, Gemini, Claude, and LLaMA, trained on vast corpora via self-supervised learning, now support increasingly multimodal tasks across text, image, and other data modalities[5,6]. The 2022 release of ChatGPT marked a major shift, expanding LLM access to broader user groups, including HEOR researchers[3,7].

Generative artificial intelligence (Gen AI), particularly large language models (LLMs), is rapidly transforming health economics and outcomes research (HEOR) by augmenting traditionally labor-intensive tasks such as systematic reviews, model development, and evidence generation[3,8]. However, the growing integration of LLMs into scientific workflows raises critical concerns around transparency, reproducibility, and trustworthiness—challenges for which HEOR-specific reporting standards do not yet exist[3,8].

In HEOR, LLMs are already being used to support systematic literature reviews (SLRs), health economic modeling (HEM), and real-world evidence (RWE) generation. These applications include tasks such as abstract screening, bias assessment, meta-analysis automation, parameter estimation, and transforming unstructured real-world data from electronic health records (EHRs), imaging, and genomics into analyzable formats [9-31]. While these uses offer substantial promise, limitations such as hallucinations, data inaccuracies, and the need for human oversight underscore the importance of structured reporting practices[3,6,8].

Regulatory and health technology assessment (HTA) bodies have begun issuing guidance. For example, the U.S. Food and Drug Administration (FDA) recently issued draft guidance proposing a risk-based credibility assessment framework for AI in regulatory submissions, including LLMs [32] and a perspective on the use of AI in its work[33]. The UK's National Institute for Health and Care Excellence (NICE) has also released both a Statement of Intent and a position statement outlining principles for generative AI use in HTA submissions [34,35], as has Canada's Drug Agency[36].

To address the lack of HEOR-specific reporting standards, the ISPOR Working Group on Generative AI developed the ELEVATE-GenAI framework. These provide structured criteria to help researchers transparently report how LLMs are used to generate or analyze evidence. While applicable for evaluation, the primary aim is to support reproducible reporting and peer review. The guidelines target studies where LLMs play a substantive role—such as in systematic reviews, economic modeling, or real-world data analysis—not those using AI for limited tasks like editing or summarization. Researchers are encouraged to apply judgment based on the context of AI use.

The article begins by presenting the literature review that informed the framework's development. Following a detailed overview of the framework and its domains, the guidelines are applied to two HEOR use cases—one in systematic review and one in economic modeling—to illustrate practical use. As a living guideline, ELEVATE-GenAI could evolve with community input and advances in generative AI. Future updates would be versioned and publicly available, with structured piloting and validation led by the ISPOR Working Group on Generative AI to ensure continued relevance, completeness and usability.

**Methods**

The ELEVATE-GenAI reporting guidelines were developed through a multi-step process involving a targeted literature review, iterative framework construction, and initial application to published HEOR use cases.

*Targeted Literature Review*

A targeted literature review was conducted to identify existing evaluation frameworks, reporting guidelines, and governance principles relevant to the use of LLMs in healthcare and health research. Searches were conducted in PubMed (through January 31, 2025) and ArXiv (through December 31, 2024), and additional reporting guidelines were retrieved from the EQUATOR Network[37], a clearing house for reporting guidelines. The search strategy, eligibility criteria, and PRISMA flow diagram are available in the Supplemental Materials. Title and abstract screening were conducted by a single reviewer (RF) using predefined eligibility criteria. Full-text screening was conducted by RF, with input from a second reviewer (JC) for uncertain cases. Data extraction was completed using a structured template to capture article title, purpose, and proposed reporting elements. Extraction was independently reviewed on a sub-sample of articles by additional co-authors (JB, JC, XW).

*Framework Development*

Findings from the targeted literature review informed the identification of key reporting domains for LLM use in HEOR. These were refined through iterative discussions within the ISPOR Working Group on Generative AI, drawing on technical literature, regulatory guidance, and real-world use cases. The framework was designed for flexibility across core HEOR applications—SLRs, HEM, and RWE—covering both high-level tasks and sub-tasks (e.g., abstract screening, model specification). To test usability and relevance, the framework was applied to two published HEOR studies: one focused on systematic review[16] and one on economic modeling[23, to assess domain coverage across different use cases.

The ELEVATE-GenAI framework is intended as a living guideline that will be refined through structured validation. Planned next steps include stakeholder consultation with researchers, industry experts, and regulatory bodies, piloting in active HEOR studies and a formal Delphi process to assess the clarity, relevance, and utility of each reporting domain. These activities, modeled on best practices from prior guideline development efforts (e.g., PRISMA-AI[38]), will support broader adoption and ensure the framework remains scientifically rigorous, usable, and adaptable as the field of generative AI evolves.

**Results**

*Literature Search Results*

A total of 522 records were identified through PubMed and ArXiv searches. After title and abstract screening, 490 records were excluded, and 32 full-text articles were assessed for eligibility. Of these, 17 were excluded, resulting in 15 studies included in the final synthesis[3,4,39-51]. An additional 6 reporting guidelines[38,52-56] and 9 position statements or frameworks[32,34,57-63] from international organizations, regulatory agencies, or HTA bodies (e.g., NICE, FDA) were included yielding a total of 30 sources included in the literature review. The Supplemental Material provides the search strategy, eligibility criteria, PRISMA flow diagram, and a table summarizing the included studies and reports.

*Overview of Literature Identified*

The 15 studies proposing evaluation frameworks included systematic reviews, conceptual models, and benchmarking protocols across domains such as clinical research, general medicine, evidence synthesis, and health technology assessment[3,4,39-51]. Nine guidance documents from national agencies, international organizations, and HTA bodies were identified[32,34,57-63]. While some focused broadly on AI/Machine Learning (ML) rather than on LLMs specifically, they were included for their relevance to responsible AI use in healthcare. Six reporting guidelines on AI and LLMs in health research were also identified [38,52-56]. These include extensions of existing standards (PRISMA-AI [38], TRIPOD+AI [53], TRIPOD-LLM[52]) as well as consensus-based checklists focused more broadly on ML (PALISADE[55], REFORMS [54]). These guidelines informed the development of ELEVATE-GenAI by highlighting principles such as model transparency, reproducibility, structured human evaluation, and ethical AI practices. In May 2025, the DEAL checklist was published and will be included in future iterations of the ELEVATE-GenAI framework[64].

*Domain identification for the ELEVATE-GenAI framework*

The ELEVATE-GenAI framework builds on domains by Bedi et al.[40] and the HELM benchmark[45], which provide strong foundations for evaluating AI performance. The ISPOR Working Group on Generative AI expanded this structure with three additional domains—Model

Characteristics; Reproducibility and Generalizability; and Security and Privacy—to address HEOR-specific methodological and regulatory needs. These additions were informed by expert input and gaps identified in the literature review. To assess alignment, components from each reviewed study were mapped to the 10 domains. **Figure 1** shows their frequency of inclusion across 30 studies, with Accuracy, Fairness and Bias, and Reproducibility and Generalizability most frequently addressed, and Security and Privacy least represented.

**Figure 1: Inclusion of ELEVATE-GenAI Domains across 30 studies and report**

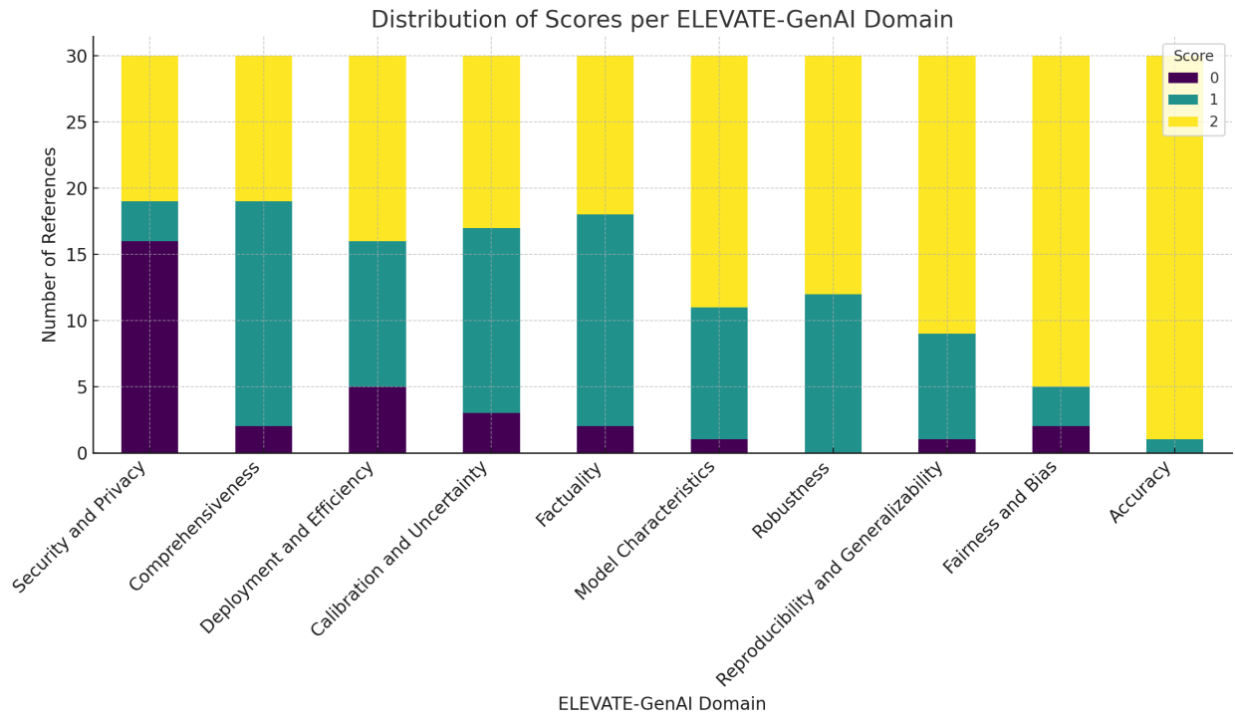

Legend: Each reference was scored across the 10 ELEVATE-GenAI reporting domains based on whether they were clearly included (Score = 2), partially included (Score = 1) or not reported (Score = 0).  The stacked bars show the number of references (N=30) receiving each score within each domain, illustrating variation in inclusion of the ELEVATE-GenAI domains across these studies.

**Reporting Domains: Definitions and Guidance**

The ELEVATE-GenAI reporting guidelines are designed for HEOR studies where generative AI plays a substantive role in evidence generation, synthesis, or analysis. They are not intended for studies using AI only for minor tasks like text editing. The 10-domain checklist covers foundational model characteristics (e.g., architecture, training data, access) and output quality across key HEOR applications such as SLRs, HEM, and RWE. Each domain includes targeted reporting items to help authors clearly describe their use of generative AI, supporting transparency and research integrity. Users should apply judgment in selecting relevant domains and briefly justify any exclusions, allowing flexibility for diverse and evolving HEOR use cases. To support interpretation, each domain is assigned a maturity level reflecting the current availability of established metrics or reporting standards. High-maturity domains have well-defined practices, while low-maturity ones indicate evolving methods. These expert-assessed ratings within the ISPOR Working Group on Generative AI are a pilot feature and will be revisited in future validation. Table 1 outlines the 10 reporting domains and their definitions.

*Model Characteristics*

This domain focuses on documenting the foundational attributes of the LLM used in the study. Key elements include the model's name (e.g., LLaMA-3), version, developer or organization, release date, license type (e.g., commercial or open source), and access method (e.g., API, web interface, or local deployment). Authors should also report the model's architecture (e.g., transformer-based) and provide details about training data sources, where applicable. This includes general-purpose pretraining corpora (where identifiable), datasets used for fine-tuning or instruction-tuning, any proprietary data used for custom models, and any sources integrated into retrieval-augmented generation (RAG) workflows. Where applicable, authors are encouraged to discuss the explainability of the model's outputs, particularly in relation to interpreting findings in HEOR contexts. While explainability is not designated as a standalone domain in ELEVATE-GenAI, it remains an important consideration for transparency, reproducibility and stakeholder trust.

**Level of maturity: High**. Well-established practices exist for describing model provenance, architecture, and access, though transparency about training data remains limited in some proprietary models.

### *Accuracy Assessment*

This domain evaluates how well Gen AI-generated outputs align with correct or expected results. Accuracy can be assessed through comparisons with human benchmarks, gold-standard datasets, or expert review. Metrics may include commonly used measures in AI/ML such as precision, recall, F1 score, and area under the curve (AUC), as well as NLP-specific (e.g., BLEU) or domain-specific metrics (e.g., GREEN for radiology report generation) [65]. In HEOR, appropriate methods include fact-checking against source documents, expert review, or benchmarking against known evidence, but the suitability of accuracy metrics depends on the task. Structured tasks like data extraction or classification lend themselves to quantitative metrics, while free-text generation—such as drafting an HTA dossier—often requires qualitative assessment. Although interest is growing in adapting AI/ML accuracy measures for HEOR tasks like SLRs and HEM, and in developing HEOR specific benchmarks, further work is needed to define fit-for-purpose evaluation strategies tailored to these specific contexts.

**Level of maturity: Medium**. Core accuracy concepts are well developed in the AI/ML field, but guidance on HEOR-specific implementation, particularly for text generation tasks, remains limited and evolving.

### *Comprehensiveness Assessment*

This domain focuses on evaluating whether GenAI-generated outputs fully and coherently address all required elements of the assigned task. In the context of HEOR, this may include ensuring that all relevant studies are captured in a systematic review, that all model components and assumptions are described in an economic evaluation, or that all relevant outcomes and perspectives are considered in value assessments. Outputs should be compared against authoritative references such as established guidelines, benchmark publications, or prior high-quality submissions. Expert review can help determine whether critical elements are missing or inadequately addressed. Comprehensiveness is distinct from accuracy: while accuracy relates to the correctness of specific elements, comprehensiveness assesses whether all relevant content has been fully and coherently addressed. For example, a meta-analysis may accurately describe included studies yet still be incomplete if it omits a pivotal trial. Ensuring completeness is essential to support informed decision-making based on the full body of evidence.

**Level of maturity: High**. While typically assessed qualitatively, there are well-established expectations for comprehensiveness across many HEOR tasks, supported by reporting guidelines and expert standards.

### *Factuality Verification*

This domain focuses on verifying that model-generated outputs are factually correct and supported by reliable sources. In HEOR, this includes confirming the accuracy of cited data, study findings, and modeling assumptions through expert review, cross-checking with primary sources, or automated source attribution where available. A key concern is the identification and correction of hallucinated or fabricated content—such as false citations, misrepresented results, or unsupported claims [19]. Authors should document any discrepancies found during review and describe the steps taken to address them. Factuality is distinct from accuracy: while accuracy reflects alignment with expected results or benchmarks, factuality concerns the truthfulness and verifiability of the content itself. For instance, a summary may accurately capture a study's structure but misreport specific findings, resulting in factual errors despite an otherwise accurate format. These distinctions, while nuanced, are important for ensuring trust in LLM-generated outputs and will be further evaluated during the piloting and validation phases described in this manuscript.

**Level of maturity: High**. Established practices for fact-checking and source validation are already in place in scientific research workflows and can be readily applied to AI-generated outputs.

### *Reproducibility Protocols and Generalizability*

This domain assesses two critical aspects of reliability: reproducibility, or the ability to replicate results, and generalizability, the applicability of methods across different contexts. Reproducibility is essential for scientific credibility and policy relevance, yet it can be difficult to achieve in generative AI due to proprietary models, frequent updates, and the stochastic nature of outputs. The dynamic nature of some generative AI systems—particularly those that continuously learn or are regularly updated—further complicates reproducibility. To mitigate these challenges, researchers should document key contextual details, including model version, date of access, deployment method (e.g., API or local instance), prompt wording, and relevant system settings (e.g., temperature, seed) [54,66]. When full transparency is not possible—especially with commercial or black-box models—authors should clearly state these limitations. Retrieval-augmented generation (RAG) approaches may enhance reproducibility by grounding model outputs in verifiable sources, providing a potential pathway for more consistent and auditable results across studies[67,68].

Generalizability involves assessing whether the LLM workflow can be applied to other HEOR questions, populations, or settings. For narrow or single-use applications, authors should indicate that generalizability does not apply and briefly explain why. Both dimensions help ensure responsible, scalable use of LLMs in HEOR.

**Level of maturity: High**. While some implementation challenges persist, particularly for closed-source systems, reproducibility documentation practices are well established, and generalizability is a routine consideration in HEOR research.

### *Robustness Checks*

This domain focuses on evaluating the model's resilience to variations in input, such as typographical errors, ambiguous phrasing, or minor changes in prompt structure. In HEOR applications, this may be particularly important for tasks that rely on consistent and interpretable output (e.g., data extraction or structured summarization). Authors should report whether robustness testing was performed and describe any observed variation in output quality or performance under perturbed input conditions. In cases where inputs and prompts are fully standardized and tightly constrained—such as in highly scripted workflows or API-based automations—robustness checks may be less relevant. Authors should briefly note when robustness testing was not conducted and explain why it was not applicable.

**Level of maturity: High**. Robustness testing is widely recognized in AI/ML research and is increasingly incorporated into evaluation practices for LLM applications in health and biomedical research.

### *Fairness and Bias*

This domain focuses on identifying and mitigating potential biases in model-generated outputs to ensure equity across populations and avoid reinforcing harmful stereotypes or exclusions. In the HEOR context, fairness may relate to how outputs differ across sociodemographic groups such as gender, age, ethnicity, or socioeconomic status. Where applicable, authors are encouraged to assess fairness using established metrics—such as demographic parity or equalized odds—and to evaluate output consistency across relevant subgroups[69-71]. However, this remains an area of active methodological development, and selecting appropriate fairness metrics and implementing subgroup analyses may require specialized expertise, particularly in HEOR applications. Authors should indicate whether fairness or bias assessments were conducted and describe any relevant findings. If this domain is not applicable to the study (e.g., if the LLM is not generating person-level or subgroup-relevant content), authors should briefly explain why it was excluded.

**Level of maturity: Low**. While fairness is a critical consideration, practical guidance and validated metrics for generative AI in HEOR remain limited and evolving.

### *Deployment Context and Efficiency Metrics*

This domain addresses both the technical configuration of the model deployment and the efficiency of its operation. Authors should describe the deployment setup, including hardware specifications (e.g., number and type of GPUs such as NVIDIA A100, H100 or TPU variants), software frameworks (e.g., Hugging Face Transformers) and orchestration tools (e.g. Docker, Ray), When possible, authors should indicate whether deployment artifacts—such as container images, configuration files, environment specifications, or API wrappers—are publicly available to facilitate reproducibility. Efficiency metrics are also essential for assessing the model's scalability and practical utility in HEOR applications. Relevant metrics may include latency (response time per query), throughput (e.g. documents processed per second), and compute efficiency (e.g. FLOPs per token) and cost metrics (e.g., token cost for commercial APIs). For example, time and cost required to generate outputs for tasks such as SLRs or HEMs may significantly influence feasibility of large-scale deployment. When models are accessed via APIs (e.g., commercial models like GPT-4o), efficiency considerations should also include token limits, response latency, usage costs, and rate limits, all of which may affect scalability, reproducibility, and real-world applicability.

**Level of maturity: High**. Clear practices exist for reporting deployment configurations and performance metrics, especially for reproducible research and cloud-based applications.

*Calibration and Uncertainty*

This domain evaluates whether the model expresses uncertainty appropriately and whether its confidence aligns with actual performance. Calibration is particularly important in HEOR, where overconfidence or under confidence in outputs can lead to misinformed decisions. Metrics such as Expected Calibration Error (ECE) [72] are being explored for HEOR use but remain underdeveloped. In systematic literature reviews (SLRs), for instance, uncertainty thresholds can help flag abstracts for manual review as part of hybrid AI–human workflows [45]. However, such metrics are not yet widely adopted in HEOR and require further validation. Authors should report whether uncertainty was assessed, how it was quantified, and whether the model's confidence appeared well-calibrated for the task. If this domain is not applicable—e.g., for tasks where confidence estimation is not used—authors should state this and provide a brief justification.

**Level of maturity: Low**. Although the concept of calibration is well defined in AI/ML, practical tools and norms for uncertainty quantification in HEOR applications remain limited and evolving.

### Security and Privacy

This domain evaluates whether appropriate safeguards are in place to protect sensitive, personal, or proprietary data used during model development or output generation. In HEOR studies that involve personal health information, clinical records, or licensed content, authors should describe relevant security protocols, including encryption methods, anonymization techniques, and access controls. Where applicable, authors should also indicate whether their work complies with data protection regulations such as GDPR or HIPAA, and describe any measures taken to protect intellectual property or copyrighted material [3]. Security and privacy protections are essential to maintaining stakeholder trust, regulatory compliance, and research integrity. If the study does not involve sensitive or proprietary data, authors may state that this domain is not applicable and provide a brief explanation.

**Level of maturity: Low**. While security and privacy principles are well established in healthcare and technology, specific implementation guidance for generative AI use in HEOR is still emerging.

*Overall Score (Optional)*

The scoring system is an optional tool to help users and reviewers assess the completeness of reporting. It is not a required domain and is not needed for framework adherence. Each domain can be rated on a three-point scale: Clearly Reported (3 points), Not Applicable (3 points), Ambiguous (2 points), or Not Reported (1 point). "Clearly Reported" indicates full adherence to domain criteria; "Not Applicable" reflects domains irrelevant to the study; "Ambiguous" refers to incomplete or unclear reporting; and "Not Reported" means relevant information is missing. The total score, calculated by summing across domains, offers a summary of reporting completeness and may support self-assessment or peer review. However, it should not be interpreted as a measure of methodological rigor. The scoring feature is optional and designed to support consistent reporting—not to grade or rank studies. Alternative approaches, such as flagging missing critical domains, will be explored in future iterations of the framework.

**Level of maturity: Low**. While scoring systems are common in reporting guidelines, their application to LLM use in HEOR is still under development and requires further testing.

## Applications of the ELEVATE-GenAI Framework to HEOR Activities

The ELEVATE-GenAI reporting framework was applied to two published HEOR use cases to illustrate its applicability: one focused on abstract screening for a systematic literature review (SLR) [16], and the other on developing a cost-effectiveness model for health economic analysis [23]. These examples, detailed in Tables 3 and 4, illustrate how the framework can be used to systematically assess the reporting of outputs augmented with LLMs across distinct HEOR tasks.

*ELEVATE-GenAI Application to a SLR Publication:*

**Table 3** shows the application of the ELEVATE-GenAI framework to evaluate the Bio-SIEVE model in the SLR study by Robinson et al.[16]. This study investigates the use of LLMs to automate title and abstract screening for SLR in the biomedical field and assesses the performance of LLMs in exclusion reasoning, (i.e., providing the rationale for excluding an abstract). The model, instruction-tuned on LLaMA and Guanaco, uses a 7B parameter architecture with quantization (4-bit LoRA) and was trained on 7,330 Cochrane systematic reviews, focusing on inclusion/exclusion criteria. Fine-tuning was validated with a curated safety-first test set to ensure task-specific performance. Accuracy metrics such as precision, recall, and overall accuracy demonstrated superior performance compared to logistic regression and other LLMs (e.g., ChatGPT). Comprehensiveness was validated against gold-standard datasets and expert reviews to ensure no relevant abstracts were missed. Factuality verification involved cross-checking inclusion/exclusion decisions with expert datasets, with discrepancies documented and addressed. Reproducibility protocols included detailed documentation of fine-tuning parameters and workflows, with publicly available code and weights for independent validation. The methods are likely generalizable to other medical domains. Robustness was assessed by varying input prompts, with Bio-SIEVE consistently excluding irrelevant abstracts. Fairness and bias monitoring were not explicitly measured. Deployment metrics, including hardware specifications (e.g., 4 A100 GPUs) and processing time (e.g., 1.39 seconds per sample), highlighted scalability. Calibration and uncertainty measures were limited, relying on manual validation without explicit thresholds for ambiguous cases. Security and privacy were addressed through anonymization and secure handling of Cochrane data, but copyright protection was not discussed. Compliance with HIPAA or GDPR would not be relevant to this type of study.

In summary, the application of Bio-SIEVE study by Robinson et al. [16] found that 6 domains were "clearly reported", 2 were "ambiguous" and 2 were "not reported". As expected, three out of the four domains that were evaluated as ambiguous or not reported (Fairness and Bias Monitoring, Calibration and Uncertainty, Security and Privacy Measures) correspond to domains with a low level of maturity for metrics, further highlighting the need for future work to identify the useful metrics for these domains.

*Application to a Health Economic Modeling Publication:*

**Table 4** demonstrates the application of the ELEVATE-GenAI framework to a health economic modeling study by Reason et al. [23]. The study explores the feasibility of using GPT-4 to automatically program health economic models. Specifically, the study aims to replicate two existing health economic analyses: the cost-effectiveness of nivolumab versus docetaxel for non-small cell lung cancer (NSCLC) and nivolumab plus ipilimumab versus sunitinib and pazopanib for renal cell carcinoma (RCC). The authors provided a detailed description of GPT-4, the LLM used in their study. Accuracy was demonstrated by replicating published three-state models (progression-free, progressed disease, and death states) with outputs aligning closely to benchmark results, as assessed by comparing incremental cost-effectiveness ratios (ICERs) to published values. For NSCLC models, 93% of runs were error-free, while RCC models required simplification but still achieved accuracy within 1% of published ICERs. Precision and recall metrics are not applicable to this use case. Comprehensiveness was validated through benchmarking and replication of complete models, though the need to simplify complex RCC calculations highlighted some limitations. Factuality verification cross-referenced ICERs and transition values with published sources, with minor discrepancies attributed to differences in discounting methods. Reproducibility was supported by detailed prompts, API parameters, and automation workflows, with generated R scripts made publicly available. Generalizability was demonstrated by the successful reuse of prompting strategies from the NSCLC model in the RCC model without modification, suggesting their potential applicability across different health economic decision problems. Robustness was tested by varying prompts, revealing limitations in handling atypical scenarios, such as overly complex calculations for RCC. Fairness was not explicitly addressed, as the study focused on technical replication rather than equity considerations. Deployment relied on Python and R scripts processed on mid-range GPUs, with generation times averaging 715 seconds for NSCLC and 956 seconds for RCC. Scalability was improved through automation workflows. Calibration and uncertainty were evaluated qualitatively, with minor ICER variability noted across runs. Security and privacy were addressed by using dummy data to replace sensitive inputs, and the authors suggested private LLM instances as a future solution to enhance security and intellectual property protections. The health economic modeling study by Reason et al. [23] effectively demonstrated the use of LLMs in cost-effectiveness modeling but omitted information required for several domains in the ELEVATE-GenAI framework. The evaluation found that 7 domains were "clearly reported", 1 was "ambiguous" and 2 were "not reported". One of the domains, Model Characteristics was evaluated as Ambiguous, but it would not be difficult for the authors in further iterations to report the appropriate information for this domain, indicating why the ELEVATE-GenAI framework has an important role to play in standardizing what authors might report.

**Limitations of the ELEVATE-GenAI Reporting Guidelines**

The ELEVATE-GenAI guidelines provide a foundational framework for reporting LLM use in HEOR, but several limitations should be acknowledged. First, the targeted literature review informing the framework was not systematic and may have omitted relevant sources. The 10 domains were derived through expert consensus and literature synthesis, but further validation is needed to ensure all relevant aspects of LLM use in HEOR are captured without introducing unnecessary complexity and reporting burden. Maturity levels for each domain reflect expert judgment and are inherently subjective; their value will need to be tested through stakeholder feedback. Similarly, while a scoring system was piloted to support self-assessment, its future utility will depend on broader user input.

Second, certain domain definitions may be challenging to apply consistently, as they are conceptually similar. For example, distinguishing between accuracy and comprehensiveness is not always straightforward—an LLM may correctly report included studies (accuracy) but fail to capture all relevant ones (comprehensiveness). Reproducibility is also difficult to achieve, given variability in data access, prompt design, and computational environments. Even with open-source models, exact replication may not be possible, and closed-source models like GPT-4 introduce further uncertainty due to ongoing updates.

Third, the framework's generalizability across HEOR tasks requires further empirical testing. While designed to be broadly applicable, it has only been applied to two use cases. As it is tested across a wider range of activities—such as SLRs, HEM, and RWE generation—its strengths and limitations will become clearer.

Fourth, many evaluation metrics commonly used in AI/ML—such as Expected Calibration Error (ECE), robustness and accuracy metrics—have not been validated for HEOR-specific tasks like parameter estimation or health state identification. Fairness and bias assessment remain particularly challenging, especially in the context of HEOR studies. Of note, benchmarks specific to HEOR field are needed. One example might be a benchmark to evaluate the accuracy of a LLM to screen titles and abstracts in a systematic literature review. To signal the variability in metric maturity, the guidelines assign a "level of maturity" to each domain. Future work should prioritize adapting these metrics to HEOR, refining reporting guidance.

Finally, as agentic approaches become more prevalent—where LLMs perform iterative or semi-autonomous tasks—future versions of ELEVATE-GenAI may require additional guidance in this area.

## Next Steps

This version of the ELEVATE-GenAI reporting guidelines was developed through expert input and a targeted literature review. Revisions to date have clarified that scoring is optional, acknowledged the absence of a standalone explainability domain, and recognized that not all domains will apply to every use case. As a living guideline, future versions will be publicly released with opportunities for community input. Next steps could include structured stakeholder consultation, piloting across a range of HEOR applications, and a formal Delphi process to assess the relevance, clarity, and utility of each domain. These activities—modeled after best practices from guideline initiatives such as PRISMA-AI[38]—will ensure the framework remains practical, flexible, and responsive to the evolving landscape of generative AI in HEOR.

## Conclusion

As the use of generative AI accelerates within HEOR, there is an urgent need for rigorous, consistent, and transparent reporting practices. LLMs offer promising capabilities to support evidence generation across tasks such as systematic literature reviews, economic modeling, and real-world data analysis. The ELEVATE-GenAI reporting guidelines provide a structured approach for documenting both model characteristics and output quality, helping to ensure scientific integrity in AI-augmented research. Initial applications of the guidelines have identified important areas for refinement, particularly around reproducibility, robustness, fairness, and uncertainty. As generative AI continues to evolve, so too must the tools used to guide its responsible integration into HEOR workflows. By adopting and iteratively improving structured reporting practices, the HEOR community can advance innovation while upholding standards of transparency and trustworthiness.

**Glossary** (adapted from Fleurence, 2024a)[3]

• **Artificial Intelligence (AI):** A broad field of computer science that aims to create intelligent machines capable of performing tasks typically requiring human intelligence.

• **Area Under the Curve (AUC):** A performance metric for classification models that measures the ability to distinguish between classes. It represents the area under the Receiver Operating Characteristic (ROC) curve, summarizing the trade-off between sensitivity (recall) and specificity. A higher AUC indicates better model performance.

• **Deep Learning:** A subset of machine learning algorithms that uses multilayered neural networks, called deep neural networks. These algorithms are the core behind the majority of advanced AI models.

• **Expected Calibration Error (ECE):** A metric that evaluates how well a model's predicted probabilities align with the actual likelihood of an event occurring. Low ECE indicates better-calibrated predictions, which is critical for applications requiring reliable confidence scores.

• **F1 Score:** A metric that balances precision and recall, calculated as the harmonic mean of these two measures. It is particularly useful for evaluating models in scenarios where false positives and false negatives have unequal consequences.

• **Foundation Model:** Large-scale pretrained models that serve a variety of purposes. These models are trained on broad data at scale and can adapt to a wide range of tasks and domains with further fine-tuning.

• **Generative AI:** AI systems capable of generating text, images, or other content based on input data, often creating new and original outputs.

• **Generative Pre-trained Transformer (GPT):** A type of large language model (LLM) based on the Transformer architecture, pre-trained on large text datasets to generate human-like language. While GPT commonly refers to OpenAI's model series (e.g., GPT-4), the term also describes a broader class of transformer-based models developed by other organizations, such as Anthropic's Claude.

• **Large Language Model (LLM):** A specific type of foundation model trained on massive text data that can recognize, summarize, translate, predict, and generate text and other content based on knowledge gained from massive datasets.

• **Machine Learning (ML):** A field of study within AI that focuses on developing algorithms that can learn from data without being explicitly programmed.

• **Multimodal AI:** An AI model that simultaneously integrates diverse data formats provided as training and prompt inputs, including images, text, bio-signals, -omics data, and more.

• **Precision:** A metric that evaluates the proportion of true positive predictions among all positive predictions made by a model. High precision indicates fewer false positives, which is essential in tasks where accuracy of positive classifications is critical.

• **Prompt:** The input given to an AI system, consisting of text or parameters that guide the AI to generate text, images, or other outputs in response.

• **Prompt Engineering:** Creating and adapting prompts (input) to instruct AI models to generate specific outputs.

• **Recall:** A metric that evaluates the proportion of true positive predictions among all actual positive cases. High recall indicates fewer false negatives, which is crucial for tasks where capturing all relevant instances is a priority.

• **Token:** A token refers to a unit of input data used by a model, which may be a word fragment, symbol, or, in the case of multimodal models, a non-text element such as an image embedding. The context window defines the maximum number of tokens a model can process at once, and determines the length and complexity of input it can handle efficiently.

**Table 1: An Evaluation Framework for Large-language models focused on Evidence, Transparency, and Efficiency (The ELEVATE-GenAI Framework) (adapted from HELM and Bedi et al.)**

| Domain Name | Domain Description | Reporting Guidelines | Level of Maturity of Domain Measurement |
|---|---|---|---|
| Model Characteristics | Describes the model's foundational characteristics, such as name, version, developer, model access, license, release date, architecture, training data, and fine-tuning performed for specific tasks. | - Provide details of the model, including name, version, developer(s), release date, license (e.g. commercial or open-source), access (e.g., links to the models), architecture (e.g., transformer-based).<br> - Describe training data, including domain-specific sources (e.g., PubMed) and any fine-tuning performed. | High |
| Accuracy Assessment | Measures how closely the model's output aligns with the correct or expected answer, evaluating precision, relevance, and correctness. | - Compare results against human benchmarks or gold-standard datasets for validation.<br>- If appropriate for the task at hand, report metrics (e.g., Precision, Recall, F1 Score, AUC). These metrics will not be applicable to all tasks. | Medium – further work required on adapting AI/ML metrics to HEOR studies and identifying appropriate metrics for specific tasks. |
| Comprehensiveness Assessment | Assesses how thoroughly the | - Evaluate completeness by comparing outputs to | High |

| | model's output addresses all aspects of the task, ensuring completeness, coherence, and critical coverage. | benchmarks, such as published reviews or models. <br> - Use expert evaluations to confirm critical elements are addressed. | |
|---|---|---|---|
| Factuality Verification | Evaluates whether the model's output is accurate and based on verifiable sources, identifying hallucinated or non-existent citations. | - Explain methods to verify factual accuracy (e.g., expert review, source validation). <br> - Document discrepancies and corrective actions taken. | High |
| Reproducibility Protocols and Generalizability | Ensures methods and outputs can be independently verified by documenting workflows, sharing code, and specifying hyperparameters. Evaluates generalizability of approach proposed | - List reproducibility protocols, including training code, query phrasing, and hyperparameters. <br> - Share workflows to facilitate independent verification. <br> - Address generalizability of methods to similar research questions | High |
| Robustness Checks | Tests the model's resilience to input variations, such as typographical errors or ambiguous queries. | - Document robustness tests, including handling of typos, adversarial inputs, or ambiguous phrasing. <br> - Report any changes in performance under these conditions. | High |
| Fairness and Bias Monitoring | Evaluates whether the model's output is equitable and free from harmful biases or stereotypes across diverse groups and contexts. | - Monitor fairness by checking for bias in outputs related to gender, age, ethnicity, or other demographics. <br> - If appropriate, use fairness metrics like demographic parity and | Low – the use of metrics to assess fairness and bias is an ongoing area of research |

| | | document corrective actions if biases are identified. | |
|---|---|---|---|
| Deployment Context and Efficiency Metrics | Examines the technical setup, resource requirements, and efficiency metrics to evaluate practical feasibility. | - Describe deployment setup, including hardware (e.g., NVIDIA A100 GPUs) and software (e.g., TensorFlow, PyTorch) and runnable deployment code (e.g., via Docker) <br> - Report efficiency metrics like processing time, scalability, and resource efficiency. | High |
| Calibration and Uncertainty | Measures how well the model conveys uncertainty in its outputs, including confidence levels and its ability to handle ambiguity appropriately. | - If appropriate for the task at hand, describe calibration methods and metrics appropriate for the task (e.g. Expected Calibration Error) <br> - Specify thresholds for flagging outputs requiring manual review (e.g. percent of abstracts included in screening in SLR) | Low – the use of metrics to evaluate calibration and uncertainty is an ongoing area of research |
| Security and Privacy Measures | Assesses adherence to security, privacy, and data protection standards and regulations, including anonymization, secure handling, and compliance with regulations like GDPR or HIPAA, if appropriate. | - Describe security protocols, such as data encryption, anonymization, and access controls. <br> - Ensure compliance with regulations like GDPR or HIPAA if appropriate <br> -Document measures to safeguard intellectual property and copyright. | Low: identifying the appropriate metrics for this domain is an ongoing area of research |
| Overall Score | Calculates an overall score for the | Assign 3 points for each domain rated as Clearly | Low: the usefulness of |

| | evaluation using the checklist | Reported, 2 points for Ambiguous, and 1 point for Not Reported. Sum the points across all domains to calculate the overall score. | this score will need to be further evaluated through feedback from the HEOR community |
| --- | --- | --- | --- |

AUC = Area under the curve; GDPR = General Data Protection Regulation; GPU = Graphics

Processing Unit; LLM = large language model; HIPAA = Health Insurance Portability and

Accountability Act;

**Table 2: ELEVATE-GenAI Checklist for Evaluating LLM Use in HEOR Research**

| Model Characteristics |
| --- |
| Is the model's name, version, developer, release date, license (e.g., open-source or commercial), and architecture described? |
| Are the training data sources (e.g., domain-specific datasets like PubMed) and fine-tuning details provided? |
| **Accuracy Assessment** |
| Are task-specific accuracy metrics (e.g., Precision, Recall, F1 Score) reported, where applicable (accounting for the fact that different metrics will be relevant for different tasks) ? |
| Are outputs validated against human benchmarks or gold-standard datasets? |
| **Comprehensiveness Assessment** |
| Are outputs compared to relevant benchmarks (e.g., published reviews, validated models) to ensure completeness? |
| Is there expert evaluation confirming all critical elements of the task are addressed? |
| **Factuality Verification** |
| Are methods for verifying the factual accuracy of outputs (e.g., cross-referencing with sources, expert review) described? |
| Are discrepancies and corrective actions documented? |
| **Reproducibility Protocols and Generalizability** |
| Are reproducibility protocols (e.g., training code, query phrasing, hyperparameters) shared? |
| Are workflows provided to support independent verification? |
| Is the generalizability of the approach and methods to similar research questions addressed? |
| **Robustness Checks** |
| Are robustness tests (e.g., handling typographical errors, ambiguous queries) documented? |
| Are changes in model performance under these conditions reported? |
| **Fairness and Bias Monitoring** |
| Are outputs evaluated for biases or stereotypes related to gender, age, ethnicity, or other demographics? |
| Are fairness metrics (e.g., demographic parity) used (if applicable) , and corrective actions for identified biases documented? |
| **Deployment Context and Efficiency Metrics** |
| Are deployment setup details (e.g., hardware, software, runnable deployment code) clearly described? |
| Are efficiency metrics (e.g., processing time, scalability, resource usage) reported? |
| **Calibration and Uncertainty** |
| Are calibration methods (e.g., Expected Calibration Error) described (if applicable) ? |
| Are thresholds for manual review of outputs (e.g., ambiguous cases flagged in systematic reviews) specified? |
| **Security and Privacy Measures** |
| Are security protocols (e.g., encryption, anonymization, access controls) documented? <br> Is compliance with regulations like GDPR or HIPAA reported, if applicable? <br> Is compliance with intellectual property and copyright law documented ? |
| **Overall Score:** Assign 3 points for each domain rated as Clearly Reported, 2 points for Ambiguous, and 1 point for Not Reported. Sum the points across all domains to calculate the overall score. |

**Table 3: Application of the ELEVATE-GenAI Checklist to a Systematic Literature Review Study (Robinson et al.)** [16]

| Checklist Questions | Domain Evaluation | Assessment |
|---|---|---|
| **1. Model Characteristics** | | **Clearly Reported** |
| Is the model's name, version, developer, release date, license (e.g., open-source or commercial), and architecture described? Are the training data sources (e.g., domain-specific datasets like PubMed) and fine-tuning details provided? | The Bio-SIEVE model is based on instruction-tuned versions of LLaMA7B and Guanaco7B, using a 7B parameter architecture with quantization (4-bit). BIO-SIEVE is not open-source, although several elements (e.g., code, parameters) are provided. The publication date is 2023. Training involved 7,330 systematic reviews from Cochrane, focusing on inclusion/exclusion criteria and reasoning for abstract exclusion. Instruction fine-tuning was conducted to improve performance on systematic review tasks. | *This item was rated as* **Clearly Reported** *because the* model name, architecture, developer, license status, training sources, and fine-tuning procedures were all described in detail, including the use of Cochrane datasets and task-specific tuning. |
| **2. Accuracy Assessment** | | **Clearly Reported** |
| Are task specific accuracy metrics (e.g., Precision, Recall, F1 Score) reported where applicable (accounting for the fact that different metrics will be relevant for different tasks)? Are outputs validated against human benchmarks or gold-standard datasets? | The paper reports precision, recall, and accuracy metrics for inclusion/exclusion tasks, comparing Bio-SIEVE's performance to baseline models (e.g., logistic regression) and other LLMs like ChatGPT. Bio-SIEVE achieved higher recall and accuracy for inclusion/exclusion but underperformed in exclusion reasoning, where ChatGPT demonstrated better results. | *This item was rated as* **Clearly Reported** *because* precision, recall, and accuracy metrics were reported and benchmarked against human labels and multiple baselines, including other LLMs. |
| **3. Comprehensiveness Assessment** | | **Clearly Reported** |
| Are outputs compared to relevant benchmarks (e.g., published reviews, validated models) to ensure completeness? Is there expert evaluation confirming all critical elements of the task are addressed? | Bio-SIEVE's outputs were validated against gold-standard datasets (e.g., Cochrane) and expert-annotated safety-first subsets. The Bio-SIEVE Guanaco7B (Single) achieved a precision of 0.85 and a recall of 0.82 on the test set, demonstrating a strong balance between minimizing false positives and capturing relevant abstracts (but performed less well on the safety-first subset). Expert validation confirmed no critical gaps | *This item was rated as* **Clearly Reported** *because* outputs were benchmarked against gold-standard datasets, and expert validation confirmed no critical gaps in inclusion coverage |
| | in inclusion, aligning with the goal of capturing all potentially relevant abstracts during screening. | |
| **4. Factuality Verification** | | **Clearly Reported** |
| Are methods for verifying the factual accuracy of outputs (e.g., cross-referencing with sources, expert review) described? Are discrepancies and corrective actions documented? | Exclusion reasoning and inclusion/exclusion decisions were cross-referenced with expert-annotated datasets. Discrepancies (e.g. missed inclusions) were documented and analyzed, with manual reviews of ambiguous cases ensuring factual reliability. | *This item was rated as* ***Clearly Reported*** *because* inclusion/exclusion outputs were compared with expert-annotated references, and discrepancies were documented and manually reviewed. |
| **5. Reproducibility Protocols and Generalizability** | | **Clearly Reported** |
| Are reproducibility protocols (e.g., training code, query phrasing, hyperparameters) shared? Are workflows provided to support independent verification? Is the generalizability of the approach and methods to similar research questions addressed? | Detailed reproducibility information includes fine-tuning parameters (e.g., batch size, learning rate), pre-processing workflows, and access to training datasets. Access to code and adapter weights is provided on HuggingFace. The approach is generalizable to abstract screening tasks in other medical domains. | *This item was rated as* ***Clearly Reported*** *because* training parameters, code, datasets, and model adapters were shared, and generalizability to other medical domains was addressed. |
| **6. Robustness Checks** | | **Clearly Reported** |
| Are robustness tests (e.g., handling typographical errors, ambiguous queries) documented? Are changes in model performance under these conditions reported? | Robustness was tested by varying input prompts and testing irrelevancy exclusions (e.g. pairing abstracts with unrelated topics). Bio-SIEVE consistently excluded irrelevant abstracts, demonstrating robustness to input variations. | *This item was rated as* ***Clearly Reported*** *because* robustness was tested by varying inputs and pairing abstracts with unrelated content, demonstrating consistent model behavior. |
| **7. Fairness and Bias Monitoring** | | **Not Reported** |
| Are outputs evaluated for biases or stereotypes related to gender, age, ethnicity, or other demographics? Are fairness metrics (e.g., demographic parity) used (if applicable) , and corrective actions for identified biases documented? | Fairness metrics, such as demographic parity, or bias in inclusion/exclusion decisions, were not explicitly assessed. Population biases were not evaluated. | *This item was rated as* ***Not Reported*** *because* no analysis of demographic or representational bias was conducted, and fairness metrics were not applied |
| **8. Deployment Context and Metrics** | | **Ambiguous** |
| Are deployment setup details (e.g., hardware, software, runnable deployment code) clearly described? Are efficiency metrics (e.g., processing time, scalability, resource usage) reported? | The Bio-SIEVE Guanaco7B models were trained on 4 NVIDIA A100 80GB GPUs for 24-40 hours, depending on the model. Inference time was reported as 1.39 seconds per sample on an RTX 3090 GPU, but context (e.g., batch size) and memory usage metrics were not provided. | *This item was rated as **Ambiguous** because* GPU usage and inference time were reported, but key efficiency metrics such as batch size, memory consumption, and scalability were not provided. |
| **9. Calibration and Uncertainty** | | **Ambiguous** |
| Is the model's uncertainty quantified and explicitly reported (if applicable)? Are thresholds for manual review of outputs (e.g., ambiguous cases flagged in systematic reviews) specified? | Confidence in inclusion/exclusion decisions was not explicitly quantified. Manual validation of safety-first decisions suggests effective uncertainty management, but explicit thresholds were not defined. | *This item was rated as **Ambiguous** because* confidence levels and thresholds for ambiguity were not quantified, although manual validation suggests some awareness of uncertainty. |
| **10. Security and Privacy Measures** | | **Not Reported** |
| Are security protocols (e.g., encryption, anonymization, access controls) documented? Is compliance with regulations like GDPR or HIPAA reported, if applicable? Is compliance with intellectual property and copyright law documented ? | Compliance with AI regulations, copyright protection, and data security were not discussed. Patient-level data was not used, minimizing direct privacy risks. | *This item was rated as **Not Reported** because* security, privacy, and regulatory compliance were not discussed, although the study avoided using identifiable patient data. |
| **Overall Score:** Assign 3 points for each domain rated as Clearly Reported, 2 points for Ambiguous, and 1 point for Not Reported. Sum the points across all domains to calculate the overall score. | | **Clearly Reported: 6, Ambiguous: 2, Not Reported: 2** Total Score = 24/30 |

GPU = Graphics Processing Unit; LLM = Large Language Model

Note: The scoring system (3 = Clearly Reported, 3= Not Applicable, 2 = Ambiguous, 1 = Not Reported) is optional and intended for self-assessment of reporting completeness only. It does not reflect methodological rigor or study quality. The scoring system will be piloted and reassessed in future validation rounds.

**Table 4: Application of the ELEVATE-GenAI Checklist to a Health Economic Modeling Study (Reason et al.)** [23]

| Checklist Questions | Evaluation | Assessment |
|---|---|---|
| **1. Model Characteristics** | | **Ambiguous** |
| Is the model's name, version, developer, release date, license (e.g., open-source or commercial), and architecture described? Are the training data sources (e.g., domain-specific datasets like PubMed) and fine-tuning details provided? | The study utilized GPT-4, a transformer-based large language model developed by OpenAI, a commercial model. Specific GPT-4 model release date was not specified. The model was accessed via API, and no specific fine-tuning for health economic modeling was reported.<br><br>GPT-4 training data includes general-purpose datasets. Explicit adaptation for health economic modeling tasks was absent. In this study, domain-specific functionality was achieved through iterative development of contextual prompts. | **This item was rated as Ambiguous because** some key elements—such as the model release date, fine-tuning details, and use of domain-specific training data—were not reported, even though general model characteristics and access method were described. |
| **2. Accuracy Assessment** | | **Clearly Reported** |
| Are task specific accuracy metrics (e.g., Precision, Recall, F1 Score) reported where applicable (accounting for the fact that different metrics will be relevant for different tasks)?? Are outputs validated against human benchmarks or gold-standard datasets? | Accuracy was assessed by comparing model outputs to the published model results. For NSCLC, 93% of runs were completely error-free; for RCC, 60% of runs required simplification but were error-free. ICERs were within 1% of published values. | This item was rated as *Clearly Reported* because model outputs were quantitatively compared against published benchmarks, and error rates and ICER deviations were clearly documented. |
| **3. Comprehensiveness Assessment** | | **Clearly Reported** |
| Are outputs compared to relevant benchmarks (e.g., published reviews, validated models) to ensure completeness? Is there expert evaluation confirming all critical elements of the task are addressed? | Outputs replicated complete three-state models, including progression-free, progressed disease, and death states. Simplification of complex RCC model steps was noted. Benchmarking against published results ensured alignment. | This item was rated as *Clearly Reported* because the outputs included all key model components and were benchmarked against complete published models, with expert interpretation noted. |
| **4. Factuality Verification** | | **Clearly Reported** |
| Are methods for verifying the factual accuracy of outputs (e.g., cross-referencing with sources, expert review) described?<br>Are discrepancies and corrective actions documented? | ICERs and transition values were cross-referenced with published models. Minor discrepancies (e.g., discounting assumptions) were documented and attributed to differences in software calculation methods. | This item was rated as *Clearly Reported* because the model outputs were cross-checked against source materials, discrepancies were noted, and explanations were provided. |
| **5. Reproducibility Protocols and Generalizability** | | **Clearly reported** |
| Are reproducibility protocols (e.g., training code, query phrasing, hyperparameters) shared?<br>Are workflows provided to support independent verification?<br>Is the generalizability of the approach methods to similar research questions addressed? | Prompts, API parameters, and Python-based automation workflows were described, enabling reproducibility. Generated R scripts are publicly available for independent validation. Prompting strategies for the NSCLC model were re-used for the RCC model without modification suggesting their potential applicability across different health economic decision problems. | This item was rated as *Clearly Reported* because detailed prompts, parameters, and automation scripts were shared, and the reuse of prompt strategies across models supported generalizability. |
| **6. Robustness Checks** | | **Clearly Reported** |
| Are robustness tests (e.g., handling typographical errors, ambiguous queries) documented?<br>Are changes in model performance under these conditions reported? | Robustness was tested through prompt variation, such as breaking scripts into multiple prompts. Simplifications were required for overly complex RCC calculations, demonstrating some limitations in handling atypical scenarios. | This item was rated as *Clearly Reported* because prompt variations were tested, and limitations in handling complex inputs were described and interpreted in context. |
| **7. Fairness and Bias Monitoring** | | **Not Reported** |
| Are outputs evaluated for biases or stereotypes related to gender, age, ethnicity, or other demographics?<br>Are fairness metrics (e.g., demographic parity) used (if applicable) , and corrective actions for identified biases documented? | The study did not explicitly address fairness or demographic bias. Outputs were focused on technical replication of published models without discussion of bias or fairness in population representation. | This item was rated as *Not Reported* because there was no assessment of fairness or bias related to demographic factors, nor any mention of mitigation strategies. |
| **8. Deployment Context and Metrics** | | **Clearly Reported** |
| Are deployment setup details (e.g., hardware, software, runnable deployment code) clearly described? | Deployment used Python and R, with scripts processed on mid-range GPUs. Average generation times were 715 seconds for the NSCLC model and 956 seconds for the RCC model. | This item was rated as *Clearly Reported* because the computational setup and processing time were described, along with the use |
| Are efficiency metrics (e.g., processing time, scalability, resource usage) reported? | Automation using Python streamlined interactions with GPT-4, improving scalability for larger datasets by reducing manual intervention. Time to create context-specific prompts was not reported. | of automation to improve scalability. |
| **9. Calibration and Uncertainty** | | **Not Reported** |
| Is the model's uncertainty quantified and explicitly reported (if applicable)? Are thresholds for manual review of outputs (e.g., ambiguous cases flagged in systematic reviews) specified? | Model outputs varied slightly across 15 runs, despite low-temperature settings. Manual quality assurance flagged errors and confirmed minor variability in ICERs. Explicit uncertainty quantification was not performed. | This item was rated as *Not Reported* because uncertainty quantification was not performed, and there were no defined thresholds or formal handling of ambiguous outputs. |
| **10. Security and Privacy Measures** | | **Clearly Reported** |
| Are security protocols (e.g., encryption, anonymization, access controls) documented? Is compliance with regulations like GDPR or HIPAA reported, if applicable? Is compliance with intellectual property and copyright law documented ? | Dummy data replaced sensitive inputs in prompts due to concerns about LLM data retention. The paper suggests private LLM instances as a future solution to address security and intellectual property concerns. | This item was rated as *Clearly Reported* because data protection strategies were described, including the use of dummy inputs and future recommendations for secure deployment. |
| **Overall Score:** Assign 3 points for each domain rated as Clearly Reported, 2 points for Ambiguous, and 1 point for Not Reported. Sum the points across all domains to calculate the overall score. | | **Clearly Reported: 7, Ambiguous: 1, Not Reported: 2** Total Score: 25/30 |

API = Application Programming Interface; ECE = Expected Calibration Error; ICER = Incremental Cost-Effectiveness Ratio; LLM = Large Language Model; NSCLC = Non-Small Cell Lung Cancer; RCC = Renal Cell Carcinoma

Note: The scoring system (3 = Clearly Reported, 3= Not Applicable, 2 = Ambiguous, 1 = Not Reported) is optional and intended for self-assessment of reporting completeness only. It does not reflect methodological rigor or study quality. The scoring system will be piloted and reassessed in future validation rounds.

<center>**Supplemental Material**</center>

**1) Targeted Literature Review**

We conducted a targeted literature review (TLR) to identify published frameworks and reporting guidelines relevant to the evaluation of output generated by large language models (LLMs) in health-related applications. The aim was to inform the development of the ELEVATE-GenAI framework by synthesizing existing literature on best practices, evaluation domains, and reporting standards specific to LLMs in healthcare, clinical, medical, and HEOR contexts.

    **a. Research question:**

What published frameworks and reporting guidelines exist for evaluating the output of large language models (LLMs) in health-related applications, and what best practices, evaluation domains, and reporting standards are recommended in the healthcare, clinical, medical, and health economics and outcomes research (HEOR) contexts?

    **b. Search Strategies**

For LLM evaluation frameworks, we searched PubMed and arXiv for peer-reviewed and preprint publications. We also included position statements and frameworks from national and international organizations, regulatory bodies, non-profit entities, and health technology assessment (HTA) agencies. These were identified through targeted website searches and supplemented by the authors' expertise and familiarity with key initiatives in the field.

For reporting guidelines, we manually searched the EQUATOR Network (Enhancing the QUAlity and Transparency Of health Research) for relevant reporting guidelines. Additional documents were identified by reviewing the reference lists of included studies.

**Table 1: Search Terms and Filters for Literature Search**

| PubMed Search Strategy: | arXiv Search Strategy: | Reporting guideline search |
|---|---|---|
| ("large language model"[Title] OR "LLM"[Title] OR "large | order: -announced_date_first; size: 200; date_range: from 2022-11-01 to 2024-12-31; | Manual review of reporting guidelines on the **EQUATOR Network** |
| language models" OR "LLMs") AND ("evaluation" OR "framework" OR "validation"[Title/Abstract]) AND (health OR clinical OR medical OR HEOR OR healthcare).

Filters: English language, human subjects, date range: November 1, 2022 – January 31, 2025. | include_cross_list: True; terms: AND title=(LLM OR large language model OR language model) AND (evaluation OR framework OR benchmarking OR holistic) AND (health OR HEOR). | (Enhancing the QUAlity and Transparency Of health Research).

Manual review of reference lists of included articles for additional relevant guidance. |

c.   **Eligibility criteria:**

We applied the following inclusion and exclusion criteria to ensure relevance to the research question.

| Category | Inclusion Criteria | Exclusion Criteria |
|---|---|---|
| **Evaluation Framework Search** | | |
| **Language** | English language only | Non-English publications |
| **Population** | Human subjects or studies applicable to human health | Studies not involving humans |
| **Publication Date** | PubMed: Jan 1, 2022 – Jan 31, 2025 <br> arXiv: Jan 1, 2022 – Dec 31, 2024 | Publications outside the specified date range |
| **Revised criteria** | Proposed or applied a **framework for evaluating LLM-generated output** in health, medicine, or health policy. <br> Described or synthesized **key evaluation domains** (e.g., factuality, robustness, bias, safety, alignment, usefulness) applicable to LLMs in health-related contexts. <br> Presented, applied, or reviewed **reporting guidelines or checklists** for studies using LLM-generated output in healthcare or biomedical research. | Focused exclusively on the **application of LLMs to clinical decision-making** (e.g., recommending a treatment or diagnosis), without discussion of output evaluation criteria or frameworks. <br> Evaluated LLMs for **disease-specific diagnostic or management tasks** (e.g., ChatGPT for detecting melanoma), with no generalizable framework or reporting guidance. <br> Assessed LLM performance on **medical licensure or board exam content**, |

| | | without proposing broader evaluative criteria. |
|---|---|---|
| **Reporting Guideline Search** | | |
| **New criteria** | Proposes or extends a **reporting guideline** for studies using LLM-generated output in health, clinical, or research contexts<br><br>Provides a **structured checklist** or **reporting framework** for AI/ML models, with relevance to evaluation, reproducibility, or transparency<br><br>Offers reporting guidance for **systematic reviews** or **evidence synthesis** involving AI-generated output | Only evaluates model performance without reporting guidance<br>No structured recommendations or checklist<br>Not applicable to AI or LLM-generated output |

d. **Prisma Diagram**

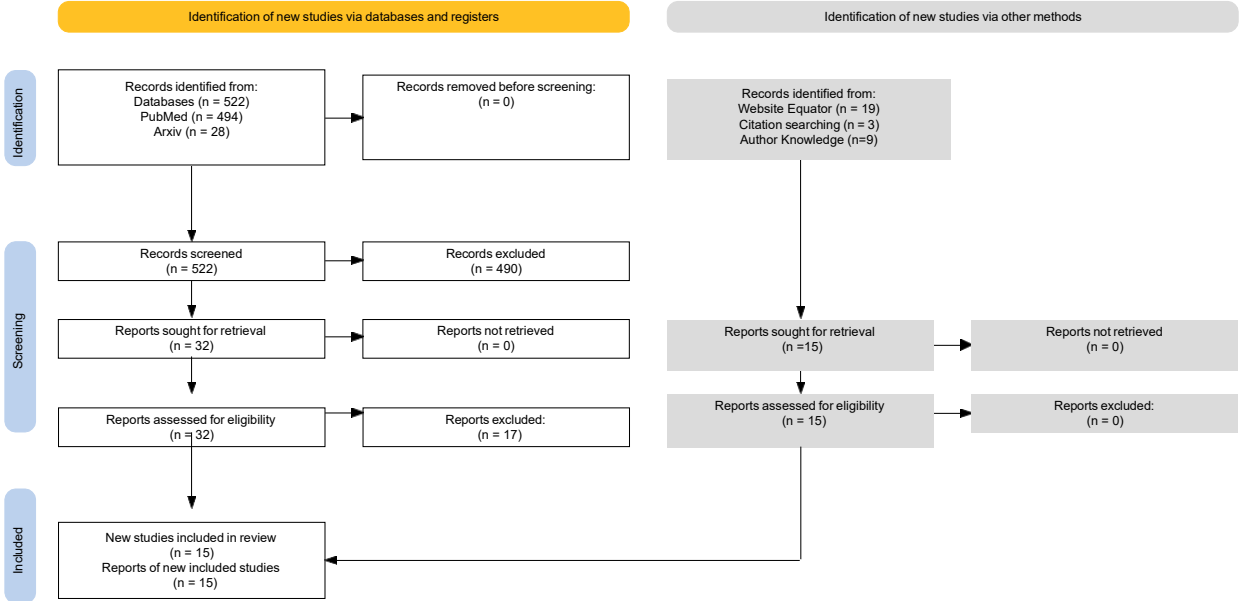

**Figure 1: PRISMA Diagram**

e.  **Results:**

Figure 2 provides a heatmap summarizing how comprehensively the 30 reviewed articles addressed each domain of the ELEVATE-GenAI framework.

**Figure 2: Heatmap of Reporting Coverage Across ELEVATE-GenAI Domains in 30 Reviewed Articles**

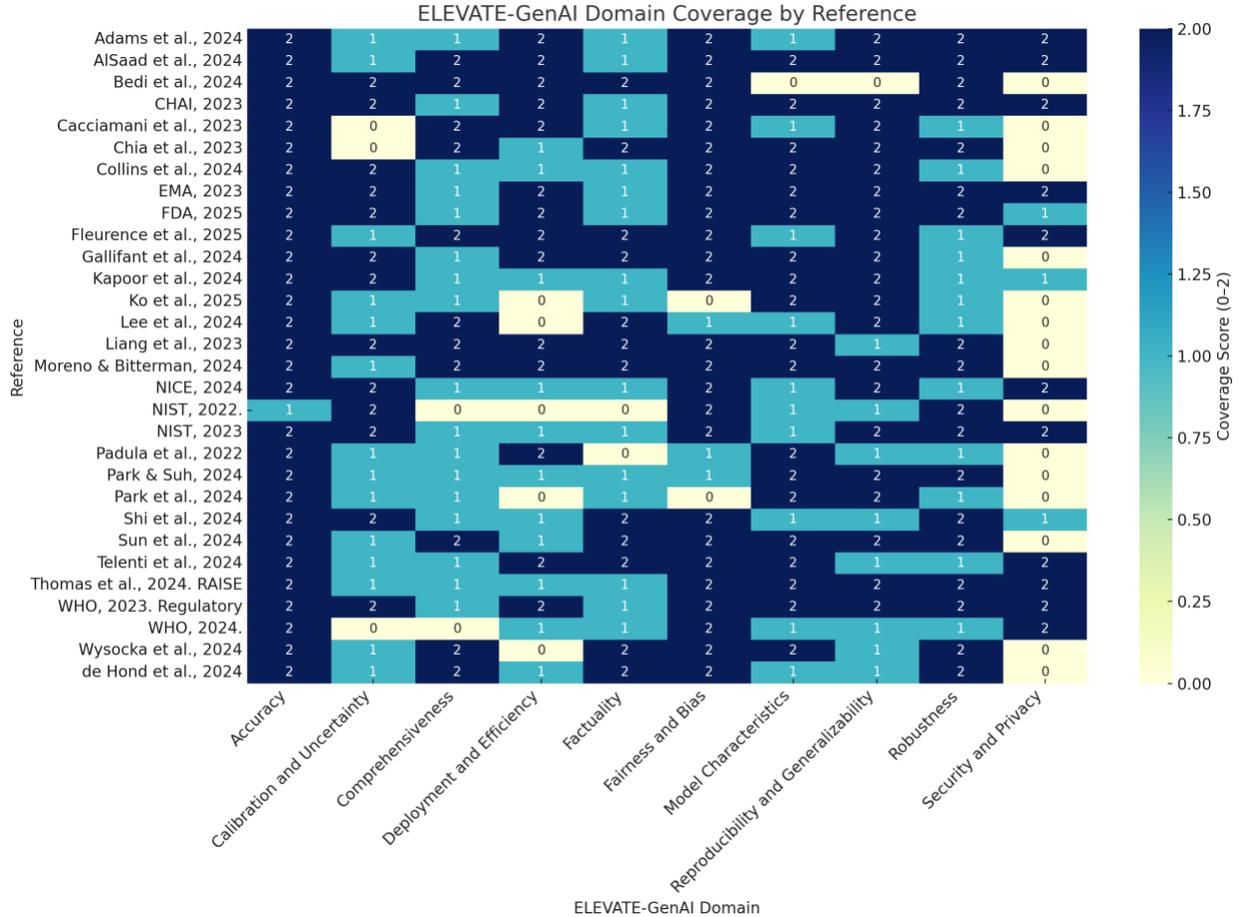

Legend: Each article was assessed against the 10 domains of the ELEVATE-GenAI reporting framework. Domains were scored as **2** (clearly reported), **1** (partially reported), or **0** (not reported).

Figure 3 shows the ELEVATE-GenAI domain coverage across the 30 included studies.

**Figure 3: Average ELEVATE-GenAI Domain Coverage across references**

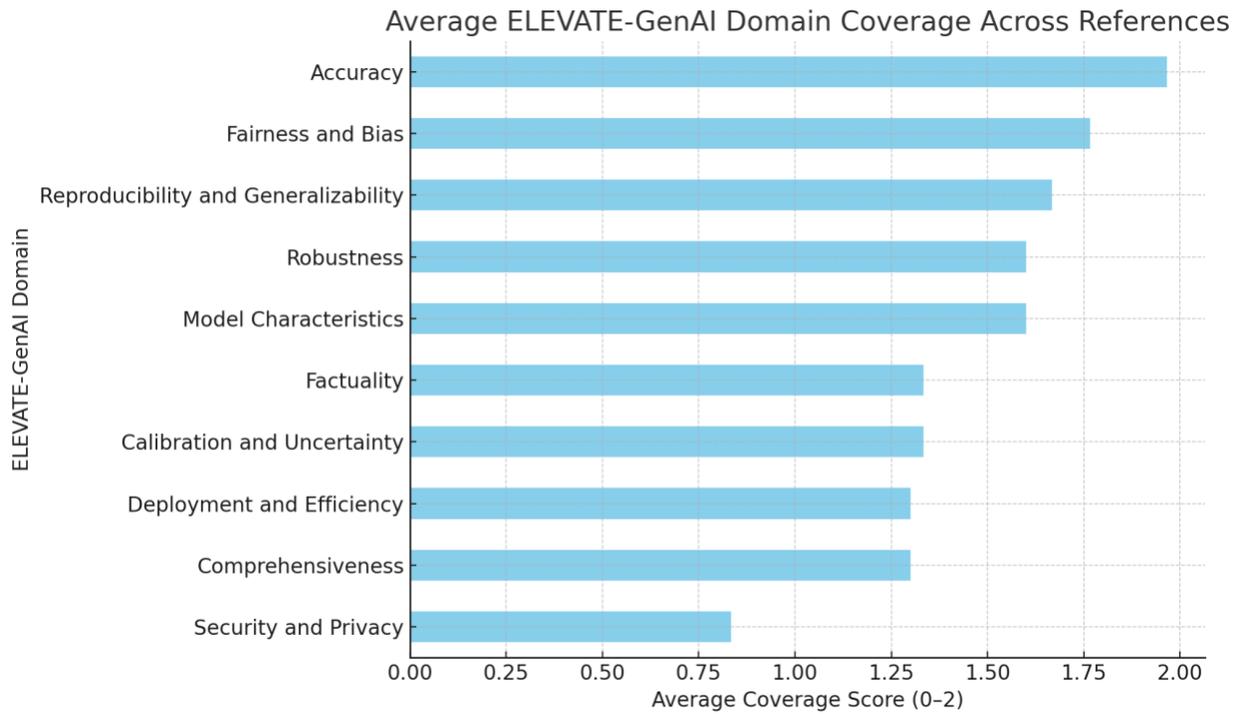

Legend: Each article was assessed against the 10 domains of the ELEVATE-GenAI reporting framework. Domains were scored as **2** (clearly reported), **1** (partially reported), or **0** (not reported).

**Table 1**: **Characteristics of Included Articles Addressing LLM Evaluation and Reporting Domains in Health Applications**

| | Reference | Title | Purpose of article | Domains proposed |
|---|---|---|---|---|
| Evaluation Frameworks | | | | |
| 1 | AlSaad et al., 2024. J Med Internet Res | Multimodal Large Language Models in Health Care: Applications, Challenges, and Future Outlook | To provide a comprehensive overview of the principles, applications, challenges, and future directions of multimodal large language models (M-LLMs) in health care. | The article identifies the need for evaluating M-LLMs based on data integration/fusion, bias and fairness, model interpretability, computational scalability, privacy and security, and alignment with clinical ethics and safety. These are framed as critical technical and ethical challenges to address when implementing LLMs in healthcare. |
| 2 | Bedi et al., 2024. JAMA | Testing and Evaluation of Health Care Applications of Large Language Models: A Systematic Review | To systematically review existing evaluations of LLMs in health care across five dimensions—data type, health care task, NLP/NLU task, dimension of evaluation, and medical specialty—and propose a structured framework for categorizing evaluation efforts. | Identifies and organizes domains used in current evaluations: accuracy, comprehensiveness, factuality, robustness, fairness/bias/toxicity, deployment metrics, and calibration/uncertainty. Recommends using real patient data, standardizing tasks and metrics, and publicly reporting failure modes. |
| 3 | Chia et al., 2023. *arXiv* | INSTRUCTEVAL: Towards Holistic Evaluation of Instruction-Tuned Large Language Models | To introduce INSTRUCTEVAL, a comprehensive human evaluation suite for instruction-tuned LLMs, evaluating performance in problem-solving, writing | Defines three core evaluation categories with specific benchmarks and rubrics: (1) **Problem-solving** (e.g., MMLU, BBH, DROP, HumanEval, CRASS), (2) **Writing ability** (IMPACT: Informative, Professional, Argumentative, Creative tasks), (3) **Alignment with human values** (HHH: Helpfulness, |

| | | | ability, and alignment to human values. | Honesty, Harmlessness). Emphasizes importance of instruction quality, task diversity, and coherence/relevance rubrics for scoring. |
|---|---|---|---|---|
| 4 | de Hond et al., 2024. *The Lancet Digital Health* | From Text to Treatment: The Crucial Role of Validation for Generative Large Language Models in Health Care | To argue for the importance of robust, multi-tiered validation processes for LLMs in healthcare, highlighting the risks of insufficient validation and the diversity of output types. | Proposes a **three-tiered validation framework**: (1) **General validation** (e.g., robustness to prompt variation, output fluency), (2) **Task-specific validation** (e.g., consistency with source data, detection of bias), and (3) **Clinical validation** (e.g., impact on patient outcomes, workflow improvements). Emphasizes the need 4for transparent reporting and integration of human evaluation in LLM validation. |
| 5 | Ko et al., 2025. *Korean J Radiol* | Adherence of Studies on Large Language Models for Medical Applications Published in Leading Medical Journals According to the MI-CLEAR-LLM Checklist | To assess how well published medical research involving LLMs adheres to the MI-CLEAR-LLM reporting checklist and to identify key gaps in transparency and reproducibility. | Evaluates adherence to the six MI-CLEAR-LLM reporting domains across 159 studies: (1) **LLM identification and specifications**, (2) **stochasticity management**, (3) **exact prompt wording and syntax**, (4) **prompt structuring**, (5) **prompt testing and optimization**, and (6) **test data independence**. Finds strong adherence to LLM identification, but major gaps in stochasticity, prompt handling, and test data reporting. |
| 6 | Fleurence et al., 2025. *Value in Health* | Generative Artificial Intelligence for Health Technology Assessment: Opportunities, | To review the applications, limitations, and policy considerations of generative AI, including large language models, in supporting key | Proposes domains important for LLM evaluation in HTA, including: (1) scientific validity and reliability, (2) bias, equity, and fairness, and (3) regulatory and ethical considerations. Also discusses reproducibility, |

| | | Challenges, and Policy Considerations | areas of health technology assessment (HTA). | transparency, and the importance of human oversight. |
|---|---|---|---|---|
| 7 | Lee et al., 2024. *BMC Med Inform Decis Mak* | Analyzing Evaluation Methods for Large Language Models in the Medical Field: A Scoping Review | To synthesize and classify evaluation methods used in LLM studies in healthcare and provide methodological guidance for future LLM evaluations. | Highlights key domains used in practice and calls for standardization. Identifies: (1) **accuracy**, (2) **concordance with expert/guideline opinion**, (3) **appropriateness**, (4) **completeness**, (5) **clarity/readability**, (6) **reproducibility**, (7) **safety/harm**, and (8) **bias and prompt transparency**. Recommends more structured evaluation designs, including repeated measurements, prompt engineering, and expert assessments. |
| 8 | Liang et al., 2023. *Trans. Mach Learn Res / arXiv* | Holistic Evaluation of Language Models | To propose HELM (Holistic Evaluation of Language Models), a comprehensive framework for evaluating LLMs across diverse scenarios and societal metrics, and to benchmark 30 prominent models under standardized conditions. | HELM proposes a taxonomy and implements evaluation across 16 core scenarios and 7 primary metrics: (1) **accuracy**, (2) **calibration**, (3) **robustness**, (4) **fairness**, (5) **bias**, (6) **toxicity**, and (7) **efficiency**. Also includes 7 targeted evaluations (e.g., knowledge, reasoning, disinformation, copyright). Emphasizes standardization, transparency, and broad scenario coverage. |
| 9 | Moreno & Bitterman, 2024. *Int J Radiat Oncol Biol Phys* | Toward Clinical-Grade Evaluation of Large Language Models | To highlight the challenges and propose rigorous strategies for pre-clinical evaluation and reproducible reporting of generative LLMs in health care, particularly for | Recommends a multi-pronged evaluation strategy including: (1) **task definition and benchmark dataset development**, (2) **transparent prompt engineering**, (3) **quantitative and human evaluation of output**, (4) **bias and fairness assessment**, and (5) **reproducibility and reporting standards**. |

| | | | cancer care and radiation oncology. | Proposes the use of expert-annotated gold-standard datasets and advocates for standardized terminology, inter-rater reliability, and alignment with clinical end-use. |
|---|---|---|---|---|
| 10 | Park et al., 2024. *Korean J Radiol* | Minimum Reporting Items for Clear Evaluation of Accuracy Reports of Large Language Models in Healthcare (MI-CLEAR-LLM) | To propose a structured checklist for transparent and replicable reporting of LLM accuracy evaluations in healthcare research. | Recommends minimum reporting domains for studies of LLMs: (1) **Identification and specifications of the LLM**, (2) **Handling of stochasticity**, (3) **Exact wording and syntax of prompts**, (4) **Detailed explanation of prompt use**, (5) **Prompt testing and optimization**, and (6) **Independence of test datasets**. The checklist aims to improve reproducibility, comparability, and rigor in studies assessing LLM output. |
| 11 | Park & Suh, 2024. *Korean J Radiol* | Reporting Guidelines for Artificial Intelligence Studies in Healthcare (for Both Conventional and Large Language Models): What's New in 2024 | To summarize recent updates to major AI reporting guidelines and highlight emerging needs for guidance tailored to studies involving large language models (LLMs). | Identifies the need for transparency in: (1) **data independence** (clarifying if test data were in the training set), (2) **prompt disclosure and usage**, (3) **management of stochasticity**, and (4) **human-AI interaction**. Recommends the upcoming CHART tool and stresses minimum standards to ensure reproducibility in LLM research. |
| 12 | Shi et al., 2024. *arXiv* | Large Language Model Safety: A Holistic Survey | To provide a comprehensive review of LLM safety across technical, ethical, and governance dimensions, and to propose a taxonomy of risks, evaluation methods, and mitigation strategies relevant | Proposes a structured taxonomy of LLM safety covering: (1) **Value misalignment** (e.g., social bias, toxicity, privacy, and ethics), (2) **Robustness to attack** (e.g., red teaming, jailbreaking, defenses), (3) **Misuse** (e.g., misinformation, deepfakes, weaponization), (4) **Autonomous AI risks** (e.g., deception, |
| | | | to LLM development and deployment. | goal misalignment), and related domains including (5) **Agent safety**, (6) **Interpretability**, (7) **Evaluation strategies**, and (8) **Governance and policy**. |
| 13 | Sun et al., 2024. *arXiv / COLM 2024* | LalaEval: A Holistic Human Evaluation Framework for Domain-Specific Large Language Models | To propose LalaEval, a comprehensive human evaluation framework for assessing domain-specific LLMs, demonstrated in the logistics industry, with standardized protocols for evaluation design, execution, and interpretation. | Proposes five major evaluation components: (1) **Domain specification**, (2) **Capability criteria (general and domain-specific)**, (3) **Benchmark dataset creation**, (4) **Evaluation rubric design**, and (5) **Systematic analysis of evaluation results**. Evaluation domains include **semantic understanding**, **factuality**, **coherence**, **creativity**, **logical reasoning**, and domain-specific capabilities such as **regulatory knowledge** and **company-specific insight**. Also incorporates rigorous grading rubrics and dispute analysis procedures. |
| 14 | Telenti et al., 2024. *Eur J Clin Invest* | Large Language Models for Science and Medicine | To review the potential applications, limitations, and broader impact of large language models in science and medicine, and propose future directions for their responsible development and use. | Identifies the need for evaluation across multiple domains, including: (1) **hallucinations and factual reliability**, (2) **bias and equity**, (3) **explainability and transparency**, (4) **validation in real-world clinical settings**, (5) **impact on decision-making and outcomes**, and (6) **regulatory, ethical, and societal implications**. Emphasizes integration with EHRs, structured evaluation designs, and the role of human oversight. |
| 15 | Wysocka et al., 2024. *J* | Large Language Models, Scientific Knowledge | To introduce and validate a framework that reduces the | Proposes a **three-step human evaluation framework**: (1) **Fluency, prompt alignment,** |
| | *Biomed Inform* | and Factuality: A Framework to Streamline Human Expert Evaluation | burden of expert evaluation in assessing LLM-generated scientific knowledge, focusing on factuality in biomedical contexts such as antibiotic discovery. | **and semantic coherence** (assessed by non-experts), (2) **Factual accuracy** (expert-reviewed), and (3) **Specificity of response**. The framework addresses **hallucinations**, **domain-specific factuality**, and **bias**, and is designed to streamline expert time while maintaining rigorous assessment. |
| **Reports from Organizations** | | | | |
| 16 | Adams et al., 2024. *NAM Perspectives* | Artificial Intelligence in Health, Health Care, and Biomedical Science: An AI Code of Conduct Principles and Commitments Discussion Draft | To present the foundational concepts and content for a harmonized draft framework–an "AI Code of Conduct"--that outlines core principles and commitments to guide the responsible development and application of AI, including LLMs, in health, health care, and biomedical science, grounded in a landscape review of existing guidelines and informed by a consensus-driven process. | **The draft framework proposes ten "Code Principles" grounded in the core values of a learning health system to promote trustworthy and responsible AI in health**: Engaged (people-centric), Safe, Effective, Equitable, Efficient (cost-effective and environmentally responsible), Accessible, Transparent, Accountable, Secure (privacy and data protection), and Adaptive (enabling continuous learning and improvement). To operationalize these values, the framework **also introduces six "Code Commitments" to apply these principles in practice**: protect and advance human health, ensure equitable distribution of benefits and risks, engage people as partners across the AI life cycle, promote workforce well-being, monitor and transparently share AI performance and impact, and continuously improve through |

| | | | | innovation and advancement of clinical practice. |
|---|---|---|---|---|
| 17 | CHAI, 2023. *Blueprint for Trustworthy AI Implementation Guidance and Assurance for Healthcare* | Blueprint for Trustworthy AI: Implementation Guidance and Assurance for Healthcare | To provide a consensus-based, practical framework to guide the implementation, evaluation, and assurance of trustworthy, safe, and effectively governed AI—including LLMs—across the healthcare ecosystem, enabling transparent and equitable adoption across stakeholders. | Proposes key domains for trustworthy AI: (1) **Usefulness** (validity, reliability, testability, usability, benefit), (2) **Safety**, (3) **Accountability and transparency** (including auditability and traceability), (4) **Explainability and interpretability**, (5) **Fairness and bias mitigation** (systemic, computational, human-cognitive), (6) **Security and resilience**, and (7) **Privacy enhancement**. Emphasizes AI lifecycle management, multidisciplinary stakeholder engagement, monitoring, and assurance infrastructure such as registries, evaluation sandboxes, and advisory services. |
| 18 | EMA, 2023. *Draft Reflection Paper on the Use of Artificial Intelligence (AI) in the Medicinal Product Lifecycle* | Reflection Paper on the Use of Artificial Intelligence (AI) in the Medicinal Product Lifecycle | To provide regulatory considerations and scientific principles for the responsible development, evaluation, and use of AI—including LLMs—across the entire lifecycle of medicinal products, from discovery through post-authorization. | Proposes domains for AI evaluation and governance: (1) **Risk-based approach** (contextual risk and regulatory impact), (2) **Data quality and acquisition**, (3) **Training, validation, and test data management**, (4) **Model development and documentation**, (5) **Performance assessment** (metrics, robustness, generalizability), (6) **Interpretability and explainability**, (7) **Deployment and monitoring**, (8) **Governance and SOPs**, (9) **Data protection and privacy**, (10) **Integrity** |

| | | | | and security, and (11) **Ethical principles** (including fairness, transparency, accountability, societal and environmental well-being, and human oversight, as per EU Trustworthy AI guidelines). |
|---|---|---|---|---|
| 19 | FDA, 2025. *Considerations for the Use of Artificial Intelligence to Support Regulatory Decision-Making for Drug and Biological Products* | Considerations for the Use of Artificial Intelligence to Support Regulatory Decision-Making for Drug and Biological Products | To provide draft recommendations for sponsors and stakeholders on establishing the credibility and risk-based evaluation of AI (including LLMs) used to generate information or data for regulatory decision-making in the drug product lifecycle. | Proposes a **risk-based credibility assessment framework** including: (1) **Defining the question of interest and context of use**, (2) **Model risk assessment** (based on influence and consequence), (3) **Detailed model and data documentation** (inputs, architecture, training, features), (4) **Model evaluation** (performance, metrics, uncertainty, independence of test data), (5) **Bias identification and mitigation**, (6) **Life cycle maintenance** (ongoing monitoring, change management), (7) **Transparency and documentation** (credibility assessment plan and report), and (8) **Early engagement with regulators**. Domains emphasize transparency, data quality, risk management, reproducibility, and accountability throughout the AI model's lifecycle. |
| 20 | NICE, 2024. Use of AI in Evidence Generation: | Use of AI in Evidence Generation: NICE Position Statement | To set out NICE's position and guidance on the appropriate, transparent, and trustworthy use of AI methods—including LLMs—for evidence generation and | **Highlights domains for evaluation and reporting of AI/LLM-generated evidence**: (1) Justification for AI use, including rationale and appropriateness relative to conventional methods; (2) Human oversight and augmentation, emphasizing that AI should |
| | Position Statement | | reporting in health technology assessment (HTA) and related evaluation programs. | support—not replace—human judgment; (3) Transparency and explainability of AI methods, including use of plain language and accessible outputs; (4) Technical and external validation to ensure plausibility and reproducibility; (5) Risk assessment and mitigation, including bias, data integrity, and cybersecurity threats (e.g., prompt injection); (6) Compliance with ethical, legal, and regulatory standards, including data protection and UK governance frameworks; and (7) Use of established reporting tools (e.g., PALISADE, TRIPOD+AI, Algorithmic Transparency Reporting Standard). Emphasizes that AI should demonstrably add value and maintain trust through transparent, accountable use. |
| 21 | NIST, 2023. *AI Risk Management Framework (AI RMF 1.0)* | Artificial Intelligence Risk Management Framework (AI RMF 1.0) | To provide a comprehensive, voluntary, and use-case agnostic framework to help organizations manage the risks associated with the design, development, deployment, and use of AI technologies and systems, promoting trustworthiness, safety, and accountability. | Identifies key characteristics that contribute to trustworthy and responsible AI: (1) **Valid and reliable**, (2) **Safe**, (3) **Secure and resilient**, (4) **Accountable and transparent**, (5) **Explainable and interpretable**, (6) **Privacy-enhanced**, (7) **Fair with harmful bias managed**. The framework is structured around four interconnected core functions: **Govern**, **Map**, **Measure**, and **Manage**—each broken down into actionable categories and subcategories covering legal and regulatory |

| | | | | compliance, organizational risk culture, human oversight, data quality, documentation, monitoring, and stakeholder engagement. |
|---|---|---|---|---|
| 22 | NIST, 2022. *SP 1270: Towards a Standard for Identifying and Managing Bias in Artificial Intelligence* | Towards a Standard for Identifying and Managing Bias in Artificial Intelligence | To introduce a preliminary, socio-technical framework and preliminary guidance for understanding, identifying, measuring, and managing bias across the full lifecycle of AI systems, including LLMs and other ML models, with a focus on building public trust and reducing harm. | Identifies three core categories of AI bias— **systemic (institutional/historical)**, **statistical/computational**, and **human/cognitive**. Provides guidance for mitigating bias at three key levels: (1) **Datasets** (representation, collection, context), (2) **Testing/Evaluation/Validation/Verification (TEVV)** (metrics, uncertainty, model selection, experimental design), and (3) **Human Factors** (participatory design, human-in-the-loop, multi-stakeholder engagement, governance). Emphasizes a **socio-technical approach**, continuous lifecycle management, transparency, documentation, and the need for organizational **governance** and multi-disciplinary evaluation. |
| 23 | WHO, 2024. *Ethics and Governance of Artificial Intelligence for Health: Guidance* | Ethics and Governance of Artificial Intelligence for Health: Guidance on Large Multi-Modal Models | To provide a comprehensive ethical and governance framework for the development, deployment, and use of large multi-modal models (including LLMs) in health, emphasizing safety, effectiveness, and equity. | **Proposes WHO consensus ethical principles for use of AI for health**: (1) protect autonomy, (2) promote human well-being, human safety, and the public interest, (3) ensure transparency, explainability, and intelligibility, (4) foster responsibility and accountability, (5) ensure inclusiveness and |

| | | | |
|---|---|---|---|
| | *on Large Multi-Modal Models* | | equity, and (6) promote AI that is responsive and sustainable. Highlights additional areas of concern, including data quality and bias, privacy and data protection, and societal and environmental impact. Offers actionable recommendations for each principle and provides governance guidance across the AI lifecycle, including development, provision, and deployment. |
| 24 | WHO, 2023. *Regulatory Considerations on Artificial Intelligence for Health* | Regulatory Considerations on Artificial Intelligence for Health | To support international dialogue and provide a resource on regulatory considerations and emerging good practices for the development, evaluation, and deployment of AI technologies in health, including LLMs. | **Highlights topics of regulatory considerations**: (1) Documentation and transparency, (2) Risk management and AI systems development lifecycle approach, (3) intended use and analytical and clinical validation, (4) data quality, (5) Privacy and data protection, and (6) engagement and collaboration. Recommends a risk-based, lifecycle approach to the development, validation, and governance of AI in health—promoting transparency, data quality, privacy, and international collaboration to ensure safe and effective deployment across diverse settings. |
| **Reporting Guidelines** | | | |

| 25 | Cacciamani et al., 2023. *Nat Med* | PRISMA-AI Reporting Guidelines for Systematic Reviews and Meta-Analyses on AI in Healthcare | To propose the development of PRISMA-AI, a consensus-based extension to PRISMA tailored to systematic reviews and meta-analyses involving AI in healthcare, aimed at improving transparency, reproducibility, and clinical relevance. | **Describes the rationale and development process for an AI-specific extension of PRISMA** for reporting systematic reviews and meta-analyses involving AI. Highlights key concerns driving the need for the guideline, including: lack of standardization, underreporting of study design and bias mitigation, limited explainability of AI systems, poor transparency in data and methods, and challenges with transparency, reproducibility and clinical applicability. Emphasizes global stakeholder engagement and use of a formal Delphi consensus process. |
| 26 | Collins et al., 2024. *BMJ* | TRIPOD+AI: Updated Guidance for Reporting Clinical Prediction Models that Use Regression or Machine Learning Methods | To provide an updated reporting checklist (TRIPOD+AI) for transparent, complete reporting of studies developing or evaluating clinical prediction models using machine learning or regression. | **TRIPOD+AI outlines 27 items (with 52 sub-items) covering**:<br><br>(1) Model development and performance evaluation, (2) Data sources, preparation, and handling, (3) Fairness, including subgroup performance and equity considerations, (4) Open science practices such as protocol registration, data/code sharing, and funding disclosure, (5) Reporting clarity and completeness across study phases, and (6) |

| | | | | Patient and public involvement in study design and dissemination. |
|---|---|---|---|---|
| | | | | The guideline emphasizes transparency, reproducibility, bias mitigation, and completeness in the reporting of prediction model studies using machine learning or regression methods—whether for model development or evaluation. |
| 27 | Gallifant et al., 2024. *Nat Med* | The TRIPOD-LLM Reporting Guideline for Studies Using Large Language Models | To introduce TRIPOD-LLM, an extension of TRIPOD+AI, offering comprehensive and modular reporting guidance tailored to the unique methodological and ethical considerations of LLM studies in healthcare. | **Presents a checklist of 19 main items and 50 sub-items for reporting studies that develop, fine-tune, prompt-engineer, or evaluate LLMs in health care.** Items span components such as: (1) LLM identification and model specifications, (2) description of training data and evaluation settings, (3) prompt engineering methods, (4) documentation of human involvement in evaluation (e.g., dual annotation), (5) reporting of transparency and fairness considerations, (6) patient and public involvement, and (7) open science practices. The guideline introduces a **modular structure** with task-specific and design-specific applicability to accommodate the diverse use cases of LLMs in biomedical research |

| 28 | Kapoor et al., 2024. *Sci Adv* | REFORMS: Consensus-Based Recommendations for Machine Learning–Based Science | To introduce a consensus-based checklist (REFORMS) for improving the transparency, reproducibility, and validity of scientific studies using machine learning, including health-related research. | Proposes a comprehensive checklist with **32 items across 8 modules**, covering: (1) **Study goals**, (2) **Computational reproducibility**, (3) **Data quality**, (4) **Data preprocessing**, (5) **Modeling decisions**, (6) **Data leakage**, (7) **Evaluation metrics and uncertainty**, and (8) **Generalizability and limitations**. Domains emphasize reporting transparency, bias detection, scientific claim validity, and reproducibility standards in ML-based science. |
|---|---|---|---|---|
| 29 | Padula et al., 2022. *Value in Health* | Machine Learning Methods in Health Economics and Outcomes Research—The PALISADE Checklist: A Good Practices Report of an ISPOR Task Force | To provide methodological guidance for the use of machine learning in HEOR and propose a structured good practice checklist to improve transparency, reproducibility, and stakeholder trust in ML-based research. | The PALISADE checklist includes 8 key domains: (1) **Purpose**, (2) **Appropriateness**, (3) **Limitations**, (4) **Implementation**, (5) **Sensitivity and specificity**, (6) **Algorithm characteristics**, (7) **Data characteristics**, and (8) **Explainability**. Focuses on improving trustworthiness and alignment with decision-maker needs. |
| 30 | Thomas et al., 2024. *RAISE Guidance, OSF Preprint* | Responsible AI in Evidence Synthesis (RAISE): Guidance and Recommendations | To provide a structured, consensus-based framework for the responsible and ethical integration of AI tools, including LLMs, in evidence synthesis processes. | **Proposes 7 core domains**: (1) Transparency (documenting AI tools and inputs), (2) Preplanning (strategic planning for AI integration), (3) Credibility (ensuring reliability and validation), (4) Ethics (bias, equity, and fairness), (5) Accountability (human oversight), (6) Compliance (with regulatory and legal standards), and (7) Evaluation (ongoing assessment of AI tool impact). |

### 2) Applying the ELEVATE-GenAI Reporting Guidelines to HEOR Activities

To demonstrate its utility, the ELEVATE-GenAI Reporting Guidelines was applied to two key tasks in HEOR: SLR abstract screening and the development of a cost-effectiveness model. These examples illustrate the framework's flexibility and its ability to guide evaluations across diverse research activities within HEOR. While these tasks highlight specific applications, the ELEVATE-GenAI Reporting Guidelines is designed to be broadly applicable to a wide range of HEOR tasks involving LLM assistance, extending beyond the examples provided.

### a. Application of ELEVATE-GenAI Reporting Guidelines to SLR Abstract Screening Task augmented with LLMs:

The Supplemental Table demonstrates the generic application of the ELEVATE-GenAI Reporting Guidelines to systematic literature review (SLR) tasks, specifically focusing on abstract screening. It provides examples of reporting requirements for each evaluation domain. While this example emphasizes abstract screening for simplicity, the framework could be equally applicable to other SLR tasks, such as full-text screening and data extraction and such applications could be the focus of future work of the ISPOR Working Group on Generative AI

The ELEVATE-GenAI Reporting Guidelines might be applied as follows. For Model Characteristics, researchers should detail the model's name, version (and version history), developer(s), training data, and any task-specific fine-tuning performed. For abstract screening, metrics such as precision, recall, and F1 score may be reported under Accuracy Assessment, with comparisons to human benchmarks or gold-standard datasets to validate performance. For many specific tasks in HEOR research, identifying appropriate metrics, adapting those commonly used in the general machine learning field, remains an ongoing area of research. The Comprehensiveness Assessment ensures that the LLM captures all relevant abstracts by comparing outputs to expert-validated gold standards, while Factuality Verification focuses on confirming the reliability of the model's inclusion/exclusion decisions through source validation. Additional domains, such as Reproducibility Protocols and Generalizability and Robustness Checks, emphasize the importance of documenting workflows, sharing code, and assessing the model's resilience to input variations. Fairness and Bias Monitoring, requires the evaluation of demographic representation in screening outputs, while Security and Privacy Measures highlight data protection and regulatory compliance, including copyright protection. Finally, practical aspects such as Deployment Context and Efficiency Metrics and Calibration and Uncertainty provide insights into resource efficiency and confidence management during screening, ensuring the framework's comprehensive applicability to SLR tasks. An overall evaluation score can be calculated as described in the table.

b. **Application of ELEVATE-GenAI Reporting Guidelines to health economic model generation augmented with LLMs:**

The Supplemental Table illustrates how the ELEVATE-GenAI Reproting Guidelines might be applied to assist with the conceptual model development for cost-effectiveness models, including generating the structure and identifying health states, by outlining specific reporting requirements for each domain. The ELEVATE-GenAI Reporting Guidelines might be applied as follows. For Model Characteristics, researchers should document details about the model, such as its name, version, developer, and training data sources, and note whether fine-tuning was conducted using published cost-effectiveness models. Accuracy Assessment involves validating the LLM's proposed health state suggestions by comparing them to gold-standard models and incorporating expert validation by health economists as a benchmark. Because accuracy metrics like precision and recall may not be applicable to this use case, future work is needed to identify metrics best suited for such applications. The Comprehensiveness Assessment ensures that the LLM's outputs address all critical health states and transitions by comparing them to established benchmarks and conducting expert reviews to identify any gaps. Factuality Verification focuses on confirming the accuracy of health state definitions and transition probabilities by cross-referencing outputs with authoritative sources such as NICE guidelines or validated cost-effectiveness models, with discrepancies documented and addressed. To support transparency, the Reproducibility Protocols domain emphasizes documenting prompts, parameters (e.g., temperature settings), and workflows used to generate the model structure, enabling independent validation. The generalizability of the model's use for other research questions should also be discussed. Robustness Checks assess whether the LLM produces consistent recommendations across different input variations, such as changing the specificity of prompts (e.g., general health state suggestions versus detailed Markov model requests). Fairness and Bias Monitoring evaluates whether health state recommendations are equitable across populations and free from demographic biases. Practical feasibility is examined under Deployment Context and Metrics, requiring descriptions of the computational setup (e.g., GPU hardware, software frameworks) and metrics like processing time or scalability for large datasets. The framework also incorporates Calibration and Uncertainty measures to assess confidence in the LLM's recommendations, identifying areas where uncertainty is flagged (e.g., ambiguous or insufficiently supported health state definitions) and providing thresholds for manual review. Metrics like ECE may not be applicable to this use case. Finally, Security and Privacy Measures ensure compliance with regulatory standards, such as GDPR and HIPAA if applicable, for example by requiring data anonymization and secure handling of sensitive or proprietary datasets. Copyright protection should also be discussed. Together, these domains provide a structured approach to evaluating the application of LLMs in cost-effectiveness modeling. An overall score can be calculated as described in the table.

**Supplemental Table: Description of the features of ELEVATE-GenAI Reporting Guidelines as relevant to (1) Systematic Literature Review Abstract Screening and (2) Conceptual Model Development for Cost-Effectiveness Analysis**

| Domain Name | Examples of what to report when using LLMs to assist with Abstract Screening in a SLR | Examples of what to report when using LLMs to assist with model structure generation and health state identification |
|---|---|---|
| Model Characteristics | -Report the model details, including its name (e.g., GPT-4), version, developer (e.g., OpenAI), release date (e.g., March 2023), and architecture (e.g., transformer-based) and license (e.g. commercial model). <br> -Describe the training data sources relevant to SLR screening tasks, such as PubMed or Cochrane abstracts. <br> -Indicate if additional fine-tuning was conducted to optimize the model for | -Report the LLM's name (e.g., GPT-4), version, developer (e.g., OpenAI), release date (e.g., March 2023), license (e.g. commercial or open-source) and architecture (e.g., transformer-based). <br> -Describe the primary training data sources. Note if the LLM was fine-tuned using high-quality, existing published models (e.g., systematic reviews of cost-effectiveness models). |
| | inclusion/exclusion criteria using RLHF or other techniques. | |
| Accuracy Assessment | -If appropriate for the task at hand, report task-specific metrics (e.g., precision, recall, F1 score, AUC) to evaluate the accuracy of outputs.<br>-Compare these metrics against human benchmarks or gold-standard datasets (e.g., Cochrane screening datasets). | -Evaluate the accuracy of the LLM's proposed model structure by comparing its health state suggestions against published gold-standard models.<br>-Evaluate the accuracy of the LLM's proposed input parameters by comparing its suggested parameter values against published gold-standard models.<br>-Include human validation by expert health economists as a key benchmark. |
| Comprehensiveness Assessment | -Evaluate whether the foundation model captures all potentially relevant abstracts during screening.<br>-Validate comprehensiveness by comparing the model's outputs to a gold-standard list of abstracts identified by domain experts or exhaustive manual review.<br>-Use benchmarks such as recall metrics to measure the percentage of relevant abstracts identified, supplemented by expert analysis to identify any critical gaps in inclusion. | -Assess the comprehensiveness of the foundation model's suggested structure and parameters for the cost-effectiveness model by comparing them to benchmarks from established published models.<br>-Incorporate expert review to identify any missing health states or input parameters critical to the research task. |
| Factuality Verification | -Describe methods for verifying factual accuracy, such as cross-checking the LLM's outputs against primary sources (e.g., PubMed).<br>-Document any discrepancies identified (e.g., hallucinated citations) and corrective actions taken (e.g., excluding non-verifiable results). | -Verify the factuality of the LLM's outputs by cross-referencing health state definitions and transition probabilities with authoritative sources, such as NICE guidelines or validated cost-effectiveness models.<br>-Document discrepancies and describe how they were resolved, if applicable. |
| Reproducibility Protocols and Generalizability | -Detail reproducibility protocols, including sharing training and preprocessing code (e.g., Python scripts for data preparation), hyperparameters (e.g., learning rate = 1e-5, batch size = 32), and validation datasets (e.g., Cochrane dataset split into 80/10/10 for training/validation/testing).<br>- Discuss generalizability of approach to other research questions. | -Provide a detailed record of the prompts and parameters (e.g., temperature settings) used to generate the cost-effectiveness model structure, including query phrasing and temperature settings.<br>-Share any reproducible workflows or code that enable independent verification of the outputs.<br>- Discuss generalizability of approach to other research questions. |
| Robustness Checks | -Describe robustness checks, such as introducing typographical errors (e.g., misspelled keywords) or ambiguous phrasing in abstracts, and report performance metrics under these conditions (e.g., F1 score drop of 5%).<br>-Include qualitative assessments of handling conflicting or ambiguous inputs. | -Test the robustness of the LLM's recommendations by altering input prompts, such as varying the specificity of the request (e.g., 'suggest health states for a Hepatitis C model' vs. 'develop a five-state Markov model for Hepatitis C').<br>-Assess whether the suggested health states remain consistent across different input variations. |
| Fairness and Bias Monitoring | -Assess demographic representation in screening outputs (e.g., stratify results by study population demographics).<br>-Use fairness metrics (e.g., demographic parity) to evaluate bias. Document corrective measures for identified imbalances (e.g., reweighting or prompt adjustments).<br>- In the absence of available metrics, provide narrative discussion of issues of fairness and bias. | -Evaluate the LLM's outputs to ensure that the recommended health states and transition probabilities are equitable across populations. For example, check whether the model suggests gender- or age-specific health states that reflect documented epidemiological data and avoid perpetuating biases.<br>- In the absence of available metrics, provide narrative discussion of issues of fairness and bias. |
| Deployment Context and Metrics | -Report the deployment setup, including hardware (e.g., NVIDIA A100 GPUs), software (e.g., Python with TensorFlow), and platforms (e.g., AWS cloud infrastructure).<br>-Include efficiency metrics such as processing speed (e.g., 1,000 abstracts screened per minute) and computational costs (e.g., GPU hours used). | -Describe the deployment setup, including hardware (e.g., NVIDIA A100 GPUs) and software frameworks (e.g., TensorFlow or PyTorch).<br>-Report efficiency metrics such as time required to generate a complete model structure (e.g., 2 minutes for a 5-state Markov model) and scalability when processing larger data inputs (e.g., recommendations for 10 different disease models). |
| Calibration and Uncertainty | -Describe methods to assess confidence in inclusion/exclusion decisions during abstract screening<br>-Specify thresholds for flagging uncertain outputs for manual review (e.g., abstracts with confidence below 70%). | -Report confidence levels for the LLM's recommendations on health state definitions.<br>-Highlight areas where uncertainty is flagged, such as cases with insufficient training data or ambiguous health state definitions, |

| Security and Privacy Measures | -Document security measures for data handling, including compliance with privacy standards (e.g., GDPR). - Report safeguards for model outputs, such as encryption and access controls, and describe steps taken to protect copyrighted or proprietary content. | -Describe privacy measures applied when using sensitive data to fine-tune the LLM, ensuring compliance with ethical and regulatory standards (e.g., de-identifying patient-level data). -If the LLM incorporates proprietary data, detail steps taken to protect intellectual property and ensure secure data handling. |
|---|---|---|
| Overall Score | Assign 3 points for each domain rated as Clearly Reported, 2 points for Ambiguous, and 1 point for Not Reported. Sum the points across all domains to calculate the overall score. | Assign 3 points for each domain rated as Clearly Reported, 2 points for Ambiguous, and 1 point for Not Reported. Sum the points across all domains to calculate the overall score. |

GDPR = General Data Protection Regulation; GPU = Graphics Processing Unit; LLM = large language model; RLHF = Reinforcement Learning from Human Feedback; SLR = Systematic Literature Review